\documentclass{ws-rv9x6}
\usepackage{ws-rv-van}             
\usepackage{graphicx}
\def\minifig#1#2#3{\parbox{#1}{{#2}\figsubcap{#3}}}
\def\ie{{\it i.e.}}
\def\eg{{\it e.g.}}
\def\etc{{\it etc}}
\def\etal{{\it et al.}}

\def\mpl{\ifmmode M_{pl}\else $M_{pl}$\fi}
\def\mpl{\ifmmode \overline M_{Pl}\else $\bar M_{Pl}$\fi}
\def\to{\rightarrow}
\def\lsim{\mathrel{\mathpalette\atversim<}}

\def\grtsim{\,\,\rlap{\raise 3pt\hbox{$>$}}{\lower 3pt\hbox{$\sim$}}\,\,}
\def\lsim{\,\,\rlap{\raise 3pt\hbox{$<$}}{\lower 3pt\hbox{$\sim$}}\,\,}
                                                                               
\makeindex
\begin{document}
                                                                                
 

\chapter[Z' Phenomenology and the LHC]{Z' Phenomenology and the LHC}

\author[T. Rizzo]{Thomas G. Rizzo}

\address{Stanford Linear Accelerator Center,\\
2575 Sand Hill Rd., Menlo Park, CA, 94025, \\
rizzo@slac.stanford.edu}

\begin{abstract}
A brief pedagogical overview of the phenomenology of Z' gauge bosons is presented. Such 
particles can arise in various electroweak extensions of the Standard Model (SM). 
We provide a quick survey of a number of Z' models, review the current 
constraints on the possible properties of a Z' and explore in detail how the LHC 
may discover and help elucidate the nature of these new particles. We provide an overview of the  
Z' studies that have been performed by both ATLAS and CMS. The role of the ILC in determining 
Z' properties is also discussed. 
\end{abstract}

\body

\section{Introduction: What is a Z' and What is It Not ?}

To an experimenter, a Z' is a resonance, which is more massive than the SM Z, observed in the Drell-Yan process 
$pp(p\bar p) \to l^+l^- +X$, where $l$=$e,\mu$ and, sometimes, $\tau$, at the LHC(or the Tevatron). To a theorist, the 
production mechanism itself tells us that this new particle is neutral, colorless and self-adjoint, 
\ie, it is its own antiparticle. However, such a new state could still be interpreted in many different ways. We may     
classify these possibilities according to the spin of the excitation, \eg, a spin-0 $\tilde \nu$ in R-parity violating 
SUSY{\cite {slepton}}, a spin-2 Kaluza-Klein(KK) excitation of the graviton as in the Randall-Sundrum(RS) 
model{\cite {Randall:1999ee,Davoudiasl:1999jd}}, 
or even a spin-1  KK excitation of a SM gauge boson from some extra dimensional model{\cite {Antoniadis:1990ew,Rizzo:1999br}} 
Another possibility for the 
spin-1 case is that this particle is the carrier of a new force, a new neutral gauge boson arising from an 
extension of the SM gauge group, \ie, a true Z', which will be our subject below{\cite {review}}. 
Given this discussion it is already clear that once a new Z'-like resonance is discovered it will first be 
necessary to measure its spin as quickly as possible to have some idea what kind of new physics we are dealing with. 
As will be discussed below this can be done rather 
easily with only a few hundred events by measuring the dilepton angular distribution in the reconstructed 
Z' rest frame. Thus, a Z' is a neutral, colorless, self-adjoint, spin-1 gauge boson that is a carrier of a new force. 
{\footnote {Distinguishing a Z' from a spin-1 KK excitation is a difficult subject beyond the scope of the present 
discussion{\cite {kks}}}}. 

Once found to be a Z', the next goal of the experimenter will be to determine as well as possible 
the couplings of this new state to the particles (mainly fermions) of the SM, \ie, to {\it identify} 
which Z' it is. As we will see there are a huge number of models which predict the existence of a 
Z'{\cite {review,pdg}}. Is this new particle one of those or is it something completely new? How does it fit into a 
larger theoretical framework?

\section{Z' Basics}

If our goal is to determine the Z' couplings to SM fermions, the first question one might ask is `How many fermionic 
couplings does a Z' have?' Since the Z' is a color singlet its couplings are color-diagonal. Thus(allowing for the 
possibility of light Dirac 
neutrinos), in general the Z' will have 24 distinct couplings-one for each of the two-component SM 
fields: $u_{L_i},d_{L_i},\nu_{L_i},e_{L_i}+(L\to R)$ with $i=1-3$ labeling the three generations. (
Of course, exotic fermions 
not present in the SM can also occur but we will ignore these for the moment.) For 
such a generic Z' these couplings are {\it non-universal}, \ie, family-dependent and this can result in dangerous 
flavor changing neutral currents(FCNC) in low-energy processes. The constraints on such beasts are 
known to be quite strong from both $K-\bar K$ and $B_{d,s}-\bar B_{d,s}$ mixing{\cite {Cheung:2006tm}} as well 
as from a large number of other low-energy processes. There FCNC are generated by fermion mixing which 
is needed to diagonalize the corresponding fermion mass matrix. As an example, consider 
schematically the Z' coupling to left-handed down-type quarks in the weak basis, \ie, 
$\bar d_{L_i}^0\eta_i d_{L_i}^0 Z'$, with $\eta_i$ being a set of coupling parameters whose different values 
would represent the generational-dependent couplings. For simplicity, 
now let $\eta_{1,2}=a$ and $\eta_3=b$ and make the unitary transformation to the physical, mass eigenstate basis, 
$d_{L_i}^0=U_{ij}d_{L_j}$. Some algebra leads to FCNC couplings of the type 
$\sim (b-a) \bar d_{L_i}U^\dagger_{i3}U_{3j}d_{L_j}Z'$. Given the existing experimental constraints, 
since we expect these mixing matrix elements to be of order those in the CKM matrix and $a,b$ to be O(1), the Z' mass 
must be huge, $\sim 100$ TeV or more, and outside the reach of the LHC. Thus unless there is some special mechanism acting 
to suppress FCNC it is highly likely that a Z' which is light enough to be observed 
at the LHC will have {\it generation-independent} couplings, \ie, now the number of couplings is reduced: 
$24 \to 8$ (or 7 if neutrinos are Majorana fields and the RH neutrinos are extremely heavy).

Further constraints on the number of independent couplings arise from several sources. First, consider 
the generator or `charge' to which the Z' couples, $T'$. Within any given model the group theory nature 
of $T'$ will be known so that one may ask if $[T',T_i]=0$, with $T_i$ being the usual SM  weak isospin 
generators of $SU(2)_L$. If the answer is in the affirmative, then all members of any 
SM representation can be labeled by a common eigenvalue of $T'$. This means that $u_L$ and $d_L$, \ie, 
$Q^T=(u,d)_L$,  as well as $\nu_L$ and $e_L$, \ie, $L^T=(\nu,e)_L$ (and dropping generation labels), 
will have identical Z' couplings so that the 
number of independent couplings is now reduced from $8 \to 6(7 \to 5)$. As we will see, this is a rather common 
occurrence in the case of garden-variety Z' which originate from extended GUT groups{\cite {review}} such as $SO(10)$ or 
$E_6$. Clearly, models which do not satisfy these conditions lead to Z' couplings which are at least partially proportional 
to the diagonal SM isospin generator itself, \ie, $T'=a T_3$ . 

In UV completed theories a further constraint on the Z' couplings arises from the requirement of 
anomaly cancellation. Anomalies can arise from one-loop fermionic triangle graphs with three external 
gauge boson legs; recall that fermions of opposite chirality contribute with opposite signs to the relevant `VVA' 
parts of such graphs. In the SM, the known fermions automatically lead to anomaly cancellation in a 
generation independent way when 
the external gauge fields are those of the SM. The existence of the Z', together with gauge invariance 
and the existence of gravity, tells us that there are 6 new graphs that must also vanish to make the theory 
renormalizable thus leading to 6 more 
constraints on the couplings of the Z'. For example, the graph with an external Z' and 2 gluons tells us that 
the sum over the colored fermion's eigenvalues of $T'$ must vanish. We can write these 6 constraints as 
(remembering to flip signs for RH fields) 
\begin{eqnarray}
\sum_{color triplets,i} T'_i=\sum_{isodoublets,i} T'_i &=&0\\ \nonumber
\sum_i Y_i^2 T_i'=\sum_i Y_i T_i'^2 &=&0\\ \nonumber
\sum_i T_i'^{3}=\sum_i T'_i&=&0\,,
\end{eqnarray}
where here we are summing over various fermion representations. These 6 constraints can be quite 
restrictive, \eg, if $T'\neq aT_{3L}+bY$, then even in the simplest Z' model, 
$\nu_R$ (not present in the SM!) must exist to allow 
for anomaly cancellation. More generally, one finds that the existence of new gauge bosons will  
also require the existence of other new, vector-like (with respect to the SM gauge group) fermions to cancel 
anomalies, something which happens automatically in the case of extended GUT groups. It is natural in 
such scenarios that the masses of these new fermions are comparable to that of the Z' itself so that  
they may also occur as decay products of the Z' thus modifying the various Z' branching fractions. If these 
modes are present then there are more coupling parameters to be determined.

\section{Z-Z' Mixing}

In a general theory the Z' and the SM Z are not true mass eigenstates due to mixing; in principle, this mixing can arise 
from two different mechanisms. 

In the case where the new gauge group G is a simple new $U(1)'$, the most general 
set of $SU(2)_L \times U(1)_Y \times U(1)'$ kinetic terms in the original weak basis (here denoted by tilded 
fields) is 
\begin{equation}
{\cal L}_K=-{1\over {4}}W_{\mu\nu}^aW^{\mu\nu}_a-{1\over {4}}\tilde B_{\mu\nu} \tilde B^{\mu\nu}
-{1\over {4}}\tilde Z'_{\mu\nu}\tilde Z'^{\mu\nu}-{{\sin \chi}\over {2}}\tilde Z'_{\mu\nu}
\tilde B^{\mu\nu}\,,
\end{equation}
where $\sin \chi$ is a parameter. Here $W_{\mu}^a$ is the usual $SU(2)_L$ gauge field while 
$\tilde B_{\mu},\tilde Z_{\mu}$ are those for $U(1)_Y$ and $U(1)'$, respectively. 
Such gauge kinetic mixing terms can be induced (if not already present) at the one-loop 
level if $Tr(T'Y)\neq 0$. Note that if G were a nonabelian group then no such mixed terms would be allowed by gauge 
invariance. In this basis the fermion couplings to the gauge fields can be schematically written as 
$\bar f(g_L T_aW^a +g_Y Y\tilde B +\tilde g_{Z'}T' \tilde Z')f$. To go to the physical basis, we make the 
linear transformations $\tilde B \to B-\tan \chi Z'$ and $\tilde Z' \to Z'/\cos \chi$ which diagonalizes 
${\cal L}_K$ and leads to the modified 
fermion couplings $\bar f[g_L T_aW^a +g_Y YB +g_{Z'}(T'+\delta Y) Z']f$ where 
$g_{Z'}=\tilde g_{Z'}/\cos \chi$ and $\delta=-g_Y \tan \chi/g_{Z'}$. Here we see that the Z' picks up an additional 
coupling proportional to the usual weak hypercharge. $\delta \neq 0$ symbolizes 
this gauge kinetic mixing{\cite {gkm}} and provides a window for its experimental observation. 
In a GUT framework, being a running parameter, $\delta(M_{GUT})=0$, but can it can  
become non-zero via RGE running at lower mass scales if the low energy sector contains matter 
in incomplete GUT representations. In most models{\cite {gkm}} where this happens, $|\delta(\sim TeV)| \leq 1/2$. 

Z-Z' mixing can also occur through the conventional Higgs-induced SSB mechanism (\ie, mass mixing) 
if the usual Higgs 
doublet(s), $H_i$(with vevs $v_{D_i}$), are {\it not} singlets under the new gauge group G. In general, the 
breaking of G requires the introduction of SM singlet Higgs fields, $S_j$(with vevs $v_{S_j}$). These 
singlet vevs should be about an order of magnitude larger than the typical doublet vevs since a Z' has 
not yet been observed. As usual the Higgs kinetic terms will generate the $W,Z$ and $Z'$ masses which 
for the neutral fields look like  
\begin{equation}
\sum_i\Big[({g_L\over {c_w}}~T_{3L}Z+g_{Z'}T'Z')v_{D_i}\Big]^2+\sum_j \Big[g_{Z'}T'v_{S_j}Z'\Big]^2\,,
\end{equation}
where $c_w=\cos \theta_W$. (Note that the massless photon has already been `removed' from this discussion.) 
The square of the first term in the first sum produces the square of the usual 
SM Z boson mass term, $\sim M_Z^2Z^2$. The square of the  last term in this sum plus the square of 
the second sum produces the corresponding Z' mass term,  
$\sim M_{Z'}^2 Z'^2$. However, the ZZ' interference piece in the first sum leads 
to Z-Z' mixing provided $T'H_i \neq 0$ for at least one $i$; note that the scale of this cross term is set by the 
doublet vevs and hence is of order $\sim M_Z^2$. 

This analysis can be summarized by noting that the interaction 
above actually generates a mass (squared) matrix in the ZZ' basis: 
\begin{eqnarray}
{\cal M}^2 & =  \left( \begin{array}{cc}
                         M_Z^2 & \beta M_Z^2  \\
                         \beta M_Z^2 & M_Z'^2 
                         \end{array}\right)\,.
\end{eqnarray}
Note that the symmetry breaking dependent parameter $\beta$,  
\begin{equation}
\beta={{4c_wg_{Z'}}\over {g_L}}\Big[\sum_i T_{3L_i}T'_iv_{D_i}^2\Big]/\sum_i v_{D_i}^2\,,
\end{equation}
can be argued to be O(1) or less on rather general grounds. Since this matrix is real, the diagonalization of 
${\cal M}^2$ proceeds via a simple 
rotation through a mixing angle $\phi$, \ie, by writing $Z=Z_1 \cos \phi -Z_2\sin \phi$, \etc, which 
yields the mass eigenstates $Z_{1,2}$ with masses $M_{1,2}$; given present data we may expect 
$r=M_1^2/M_2^2 \leq 0.01-0.02$. $Z_1\simeq Z$ is the state presently produced at colliders, \ie, $M_1=91.1875\pm 0.0021$ 
GeV, and thus we might also expect that $\phi$ must be quite small for the SM to work as well as it does. 
Defining $\rho=M_Z^2/M_1^2$, with $M_Z$ being the would-be mass of the Z if no mixing occurred, we can 
approximate 
\begin{eqnarray}
\phi &=&-\beta r[1+(1+\beta^2)r +O(r^2)]\\ \nonumber
\delta \rho &=& \beta^2 r[1+(1+2\beta^2)r +O(r^2)]\,,
\end{eqnarray}
where $\delta \rho=\rho-1$, so that $\beta$ determines the sign of $\phi$. We thus expect that both 
$\delta \rho,|\phi| < 10^{-2}$. In fact, if we are {\it not} dealing with issues associated with 
precision measurements{\cite {LEPEWWG}} 
then Z-Z' mixing is expected to be so small that it can be safely neglected. 

It is important to note that non-zero mixing modifies the predicted SM Z couplings to 
${g_L\over {c_w}}(T_{3L}-x_WQ)c_\phi+g_{Z'}T's_\phi$, where $x_W=\sin^2 \theta_W$, which can lead to many  
important effects. For example, the partial width for $Z_1\to f\bar f$ to lowest order(\ie, apart from phase space, 
QCD and QED radiative corrections) is now given by
\begin{equation} 
\Gamma(Z_1\to f\bar f)=N_c {{\rho G_FM_1^3(v_{eff}^2+a_{eff}^2)}\over {6{\sqrt 2}\pi}}\,, 
\end{equation}
where $N_c$ is a color factor, $\rho$ is given above and 
\begin{eqnarray}
v_{eff}&=&(T_{3L}-2x_WQ)c_\phi+{g_{Z'}\over {g_L/(2c_w)}}(T'_L+T'_R)s_\phi\\ \nonumber
a_{eff}&=&T_{3L}c_\phi+{g_{Z'}\over {g_L/(2c_w)}}(T'_L-T'_R)s_\phi\,,
\end{eqnarray}
and where $T'_{L,R}$ are the eigenvalues of $T'$ for $f_{L,R}$. Other effects that can occur include decay modes such 
as $Z_2 \to W^+W^-,Z_1H_i$, where $H_i$ is a light Higgs, which are now induced via mixing. If $T'$ has no $T_3$ 
component this is the only way such decays can occur at tree level. In the case of the $Z_2 \to W^+W^-$ 
mode, an interesting cancellation occurs: the partial width scales as $s_\phi^2 (M_2/M_W)^4$, where 
the second factor follows from the Goldstone Boson Equivalence Theorem{\cite {equiv}}. However, 
since $s_\phi \simeq -\beta r$ and 
$r=M_1^2/M_2^2 \simeq M_Z^2/M_2^2$, we find instead that the partial width goes as $\sim \beta^2$ without any 
additional mass enhancement or suppression factors.  The tiny mixing angle induced by small $r$ has been offset by the 
large $M_2/M_W$ ratio! In specific models, one finds that this small Z-Z' mixing leads to 
$Z_2 \to W^+W^-$ partial widths which can be comparable to other decay modes. Of course, $Z_2 \to W^+W^-$ can be also be 
induced at the one-loop level but there the amplitude will be suppressed by the corresponding loop factor as well as 
possible small mass ratios.

\section{Some Sample Z' Models}

There are many (hundreds of) models on the market which predict a Z' falling into two rather broad categories depending 
on whether or not they arise in a GUT scenario. The list below is {\it only} meant to be representative and is 
very far from exhaustive and I beg pardon if your favorite model is not represented. 

The two most popular GUT scenarios are the Left Right Symmetric Model(LRM){\cite {LRM}} and those that come from 
$E_6$ grand unification{\cite {review}}. 

($i$) In the $E_6$ case one imagines a symmetry breaking pattern 
$E_6 \to SO(10)\times U(1)_\psi \to SU(5)\times U(1)_\chi \times U(1)_\psi$. Then $SU(5)$ breaks to the SM and 
only one linear combination $G=U(1)_\theta= c_\theta U(1)_\psi -s_\theta U(1)_\chi$ remains light at the TeV scale. 
$\theta$ is treated as a free parameter{\footnote {The reader should be aware that there are several different 
definitions of this mixing angle in the literature, \ie, $Z'=Z_\chi \cos \beta+z_\psi \sin \beta$ occurs quite commonly.}} 
and the particular values $\theta=0, ~-90^o, ~\sin^{-1}\sqrt{(3/8)} \simeq  
37.76^o$ and $-\sin^{-1}\sqrt{(5/8)} \simeq -52.24^o$, correspond to `special' models called $\psi, \chi, \eta$ and $I$, 
respectively. These models are sometimes referred to in the literature as effective rank-5 models(ER5M). 
In this case, neglecting possible kinetic mixing, 
\begin{equation}
g_{Z'}T'=\lambda {g_L\over {c_w}}\sqrt{{5x_W}\over {3}}({{Q_\psi c_\theta}\over{2\sqrt 6}}-
{{Q_\chi s_\theta}\over{2\sqrt {10}}})\,,
\end{equation}
where $\lambda \simeq 1$ arises from RGE evolution. The parameters 
$Q_{\psi,\chi}$ originate from the embeddings of the SM fermions into the fundamental {\bf 27} representation of 
$E_6$. A detailed list of their values can be found in the second paper in{\cite {review}} 
with an abbreviated version given in the Table 
below in LH field notation.  Note that this is the {\it standard} form for this embedding and there are other 
possibilities{\cite {review}}. These other choices can be recovered by a shift in the parameter $\theta$. 
Note further that in addition to the SM fermions plus the RH neutrino, 
$E_6$ predicts, per generation, an additional neutral singlet, $S^c$, along with an electric charge 
$Q=-1/3$, color triplet, vector-like isosinglet, $h$, and a color singlet, vector-like isodoublet whose top member has 
$Q=0$, $H$ (along with their conjugate fields). These exotic fermions with masses 
comparable to the Z' cancel the anomalies in the theory and can lead to interesting new phenomenology{\cite {review}} but we 
will generally ignore them in our discussion below. In many cases these states are quite heavy and thus will not participate 
in Z' decays. 
\begin{table}[ht]
\tbl{Quantum numbers for various SM and exotic fermions in LH notation in $E_6$ models}
{\begin{tabular}{ccc} \toprule
Representation & $Q_\psi$ & $Q_\chi$ \\
\colrule
$Q$ &  1 & -1 \\
$L$ &  1 & 3 \\
$u^c$ & 1 & -1 \\
$d^c$ & 1 & 3 \\
$e^c$ & 1 & -1 \\
$\nu^c$ & 1 & -5\\ 
$H$ & -2 & -2\\
$H^c$& -2 & 2\\
$h$ & -2 & 2\\
$h^c$& -2 & -2\\
$S^c$ & 4 & 0\\ \botrule
\end{tabular}}
\label{tbl1}
\end{table}

($ii$) The LRM, based on the low-energy gauge group $SU(2)_L\times SU(2)_R \times U(1)_{B-L}$, can arise from 
an $SO(10)$ or $E_6$ GUT. Unlike the case 
of ER5M, not only is there a $Z'$ but there is also a new charged $W_R^\pm$ gauge boson since here $G=SU(2)$. 
In general $\kappa=g_R/g_L \neq 1$ is a free parameter but must be $>x_W/(1-x_W)$ for the existence of 
real gauge couplings. On occasions, the parameter $\alpha_{LR}=\sqrt {c_w^2\kappa^2/x_W^2-1}$ is also often used. 
In this case we find that  
\begin{equation}
g_{Z'}T'={g_L\over {c_w}}[\kappa^2-(1+\kappa^2)x_W]^{-1/2}[x_WT_{3L}+\kappa^2(1-x_W)T_{3R}-x_WQ]\,.
\end{equation}
The mass ratio of the W' and Z' is given by 
\begin{equation}
{{M_Z'^2}\over {M_{W'}^2}}={{\kappa^2(1-x_W)\rho_R}\over {\kappa^2(1-x_W)-x_W}}>1\,,
\end{equation}
with the values $\rho_R=1(2)$ depending upon whether $SU(2)_R$ is broken by either Higgs 
doublets(or by triplets). The existence of a $W'=W_R$ with the correct mass ratio to the Z' provides a 
good test of this model. Note that due to the LR symmetry we need not introduce additional fermions in this model 
to cancel anomalies although right-handed neutrinos are present automatically. In the $E_6$ case a variant of 
this model{\cite {ALRM}} can be constructed by altering the embeddings of the 
SM and exotic fermions into the ordinary {\bf 10} and {\bf 5} representations (called the Alternative LRM, \ie,  ALRM).  

($iii$) The Z' in the Little Higgs scenario{\cite {Little}} provides the best non-GUT example. The new particles in these 
models, \ie, new gauge bosons, fermions and Higgs, are necessary to remove at one-loop the quadratic divergence of 
the SM Higgs mass and their natures are dictated by the detailed group structure of the particular model. 
This greatly restricts the possible couplings of such states. With a $W'$ which is essentially 
degenerate in mass with the Z', the Z' is found to couple like $g_{Z'}T'=(g_L/2)T_{3L}\cot \theta_H$, with $\theta_H$ another 
mixing parameter. 

($iv$) Another non-GUT example{\cite {Lynch:2000md}} is based on the group $SU(2)_l\times SU(2)_h \times U(1)_Y$ with $l,h$ 
referring to `light' and `heavy'. The first 2 generations couple to $SU(2)_l$ while the third couples to $SU(2)_h$. 
In this case the Z' and W' are again found to be 
degenerate and the Z' couples to $g_{Z'}T'=g_L[\cot \Phi T_{3l}-\tan \Phi T_{3h}]$ 
with $\Phi$ another mixing angle. Such a model is a good example of where the Z' couplings are generation 
dependent. 

($v$) A final example is a Z' that has couplings which are exactly the same as those of the SM Z (SSM), but is just 
heavier. This is not a real model but is very commonly used as a `standard candle' in experimental Z' searches. A 
more realistic variant of this model is one in which a Z' has {\it no} couplings to SM fermions in the weak basis 
but the couplings are then induced in the mass eigenstate basis Z-Z' via mixing. In this case the relevant couplings of the Z' 
are those of the SM Z but scaled down by a factor of $\sin \phi$. 

A nice way to consider rather broad classes of Z' models has recently been described by Carena \etal{\cite {Carena:2004xs}}. 
In this approach one first augments the SM fermion spectrum by adding to it a pair of vector-like (with respect to the SM) 
fermions, one transforming like $L$ and the other like $d^c$; this is essentially what happens in the $E_6$ GUT model. 
The authors then look for families of models that satisfy the six anomaly constraints with generation-independent 
couplings. Such an analysis yields several sets of 1-parameter solutions for the generator $T'$ but leaves the coupling 
$g_{Z'}$ free. The simplest such solution is $T'=B-xL$, with $x$ a free par meter. Some other solutions include 
$T'=Q+xu_R$ (\ie, $T'(Q)=1/3$ and $T'(u_R)=x/3$ and all others fixed by anomaly cancellation), $T'=d_R-xu_R$ and 
$T'=10+x\bar 5$, where `10' and $\bar 5$ refer to $SU(5)$ GUT assignments.

\section{What Do We Know Now? Present Z' Constraints}

Z' searches are of two kinds: indirect and direct. Important constraints arise from both sources at the 
present moment though this is likely to change radically in the near future. 

\subsection{Indirect Z' Searches}

In this case one looks for deviations from the SM that might be associated with the existence of a Z'; this 
usually involves precision electroweak measurements at, below and {\it above} the Z-pole. The cross section and forward 
backward asymmetry, $A_{FB}$, measurements at LEPII take place at high center of mass energies which are still (far) 
below the actual Z' mass.

Since such constraints are indirect, one can generalize from the case of a new Z' and consider a more encompassing 
framework based on contact interactions{\cite {contact}}. Here one `integrates out' the new 
physics (since we assume we are at 
energies below which the new physics is directly manifest) and express its influence via higher-dimensional 
(usually dim-6) operators. For example, in the dim-6 case, for the process $e^+e^- \to \bar ff$, we can consider an 
effective Lagrangian of the form{\cite {contact}} 
\begin{equation}
{\cal L}={\cal L}_{SM}+{{4\pi}\over {\Lambda^2(1+\delta_{ef})}}\sum_{ij=L,R} \eta_{ij}^f 
(\bar e_i\gamma_\mu e_i)(\bar f_j\gamma^\mu f_j)\,,
\end{equation}
where $\Lambda$ is called `the compositeness scale' for historic reasons, $\delta_{ef}$ takes care of the statistics in the 
case of Bhabha scattering, and the $\eta$'s are chirality structure 
coefficients which are of order unity. The exchange of many new states can be described in this way and can be 
analyzed simultaneously. The corresponding parameter bounds can then be interpreted within your favorite model. 
This prescription can be used for data at all energies as long as these energies are far below $\Lambda$. 

\begin{figure}[ht]
\begin{center}
  \minifig{2.2in}{\epsfig{figure=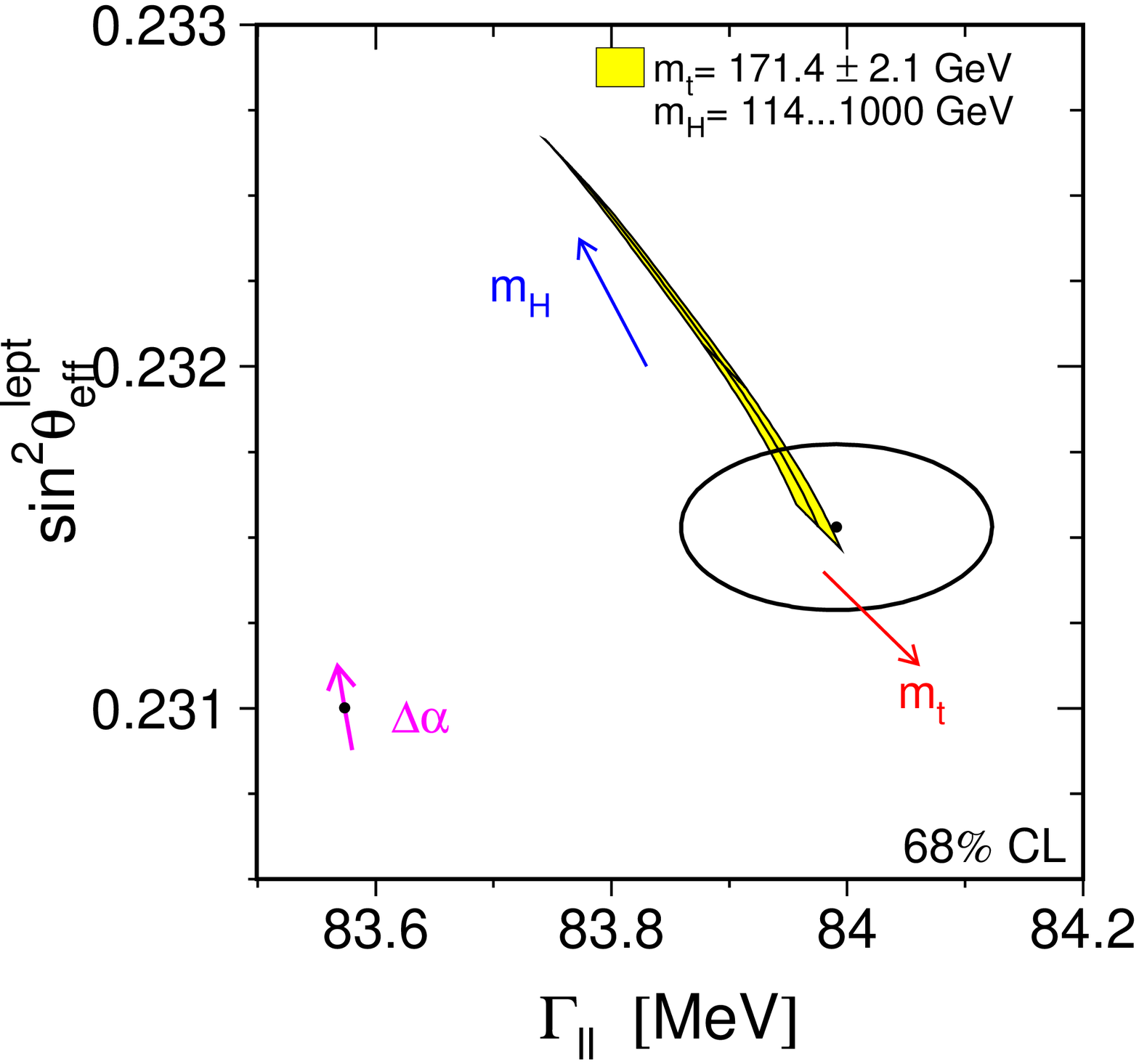,width=2.3in}}{a}
  \hspace*{4pt}
  \minifig{2.1in}{\epsfig{figure=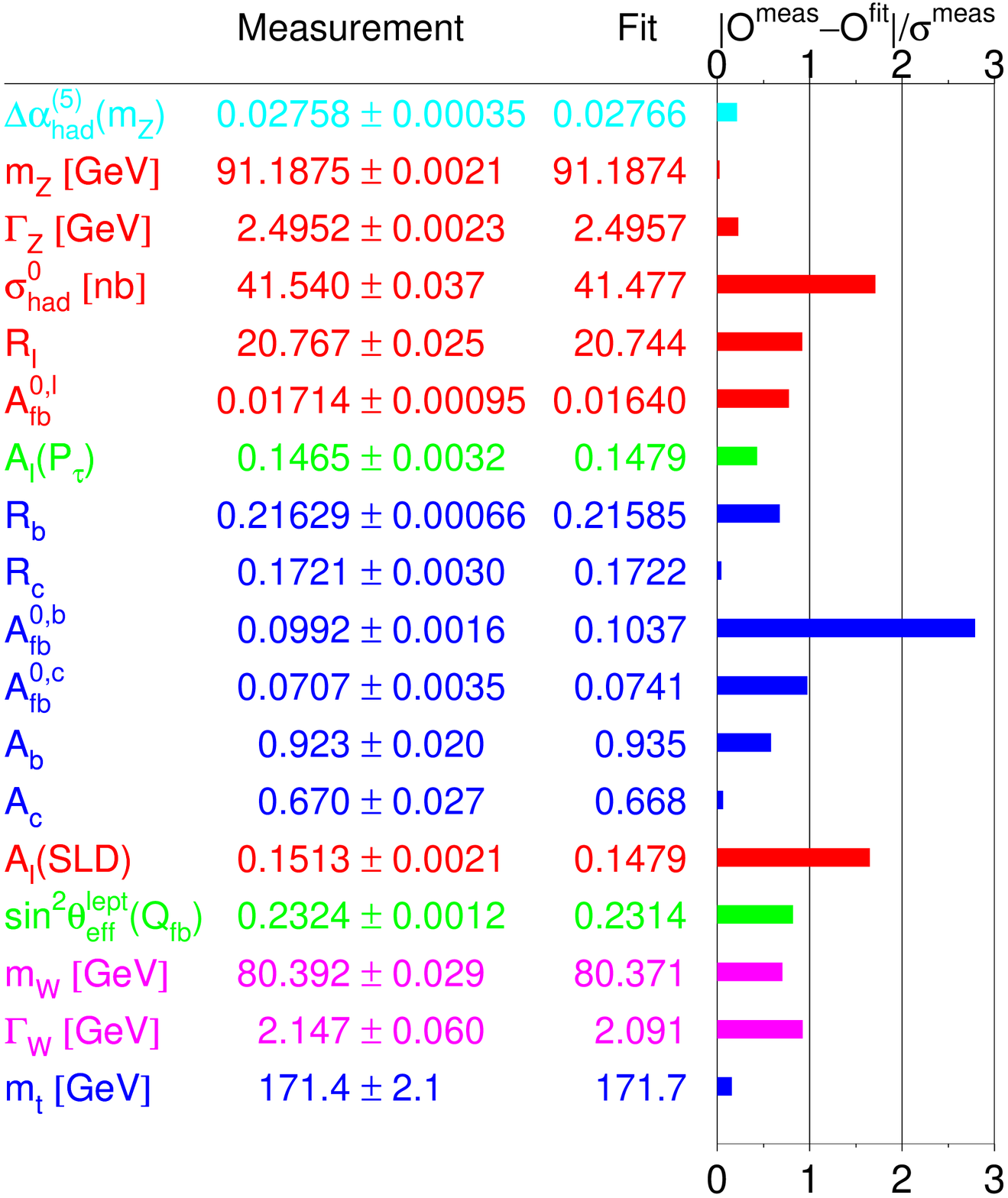,width=2in}}{b}
  \caption{Summer 2006 results from the LEPEWWG. (a) Fit for the Z leptonic partial width and $\sin^2 \theta_{lepton}$ 
in comparison to the SM prediction in the yellow band.(b) Comparison of a number 
of electroweak measurements with their SM fitted values.}%
  \label{fig1}%
\end{center}
\end{figure}

Z-pole measurements mainly restrict the Z-Z' mixing angle as they are sensitive to small mixing-induced deviations in the SM 
couplings and not to the Z' mass. LEP and SLD have made very precise measurements of these couplings which can be 
compared to SM predictions including radiative corrections{\cite {LEPEWWG}}. An example of this is found in Fig.~\ref{fig1} 
where we see the experimental results for the leptonic partial width of the Z as well as $\sin^2 \theta_{lepton}$ in
comparison with the corresponding SM predictions. Deviations in  $\sin^2 \theta_{lepton}$ are particularly sensitive 
to shifts in the Z couplings due to non-zero values of $\phi$. Semiquantitatively these measurements strongly 
suggest that $|\phi| \leq a~few ~10^{-3}$, at most, in most Z' models assuming a light Higgs. Performing a global 
fit to the full electroweak data set, as given, \eg, by the LEPEWWG{\cite {LEPEWWG}} gives comparable 
constraints{\cite {pdg}}. 

Above the Z pole, LEPII data provides strong constraints on Z' couplings and masses but are generally insensitive to 
small Z-Z' mixing. Writing the couplings as 
$\sum_i \bar f\gamma_\mu(v_{f_i}-a_{f_i}\gamma_5)fZ^\mu_i$ for $i=\gamma,Z,Z'$, the differential cross section 
for $e^+e^- \to \bar f f$ when $m_f=0$ is just 
\begin{equation}
{{d\sigma}\over {dz}}={N_c\over {32\pi s}}\sum_{i,j}P_{ij}[B_{ij}(1+z^2)+2C_{ij}z]\,, 
\end{equation}
where 
\begin{eqnarray}
B_{ij}&=&(v_iv_j+a_ia_j)_e(v_iv_j+a_ia_j)_f\\ \nonumber 
C_{ij}&=&(v_ia_j+a_iv_j)_e(v_ia_j+a_iv_j)_f\,,
\end{eqnarray}
and 
\begin{equation}
P_{ij}=s^2{{(s-M_i^2)(s-M_j^2)+\Gamma_i\Gamma_jM_iM_j}\over {[(s-M_i^2)^2+\Gamma_i^2M_i^2][i\to j]}}\,,
\end{equation}
with $\sqrt s$ the collision energy, $\Gamma_i$ being the total widths of the exchanged particles and 
$z=\cos \theta$, the scattering angle in the CM frame. $A_{FB}$ for any final state fermion $f$ 
is then given by the ratio of integrals 
\begin{equation}
A_{FB}^f= \Bigg[{{\int_0^1 ~dz {{d\sigma}\over {dz}}-\int_{-1}^0 ~dz 
{{d\sigma}\over {dz}}}\over {~~''~~+~~''~~}}\Bigg]\,.
\end{equation}
If the $e^\pm$ beams are polarized (as at the ILC but not at LEP) one can also define the left-right polarization 
asymmetry, $A_{LR}^f$; to this end we let 
\begin{eqnarray}
B_{ij}&\to& B_{ij}+\xi (v_ia_j+a_iv_j)_e(v_iv_j+a_ia_j)_f\\ \nonumber 
C_{ij}&\to& C_{ij}+\xi (v_iv_j+a_ia_j)_e(v_ia_j+a_iv_j)_f\,,
\end{eqnarray}
and then form the ratio 
\begin{equation}
A_{LR}^f(z)=P\Bigg[{{d\sigma(\xi=+1)-d\sigma(\xi=-1)}\over {~~~~''~~~~+~~~~''~~~~}}\Bigg]\,, 
\end{equation}
where $P$ is the effective beam polarization. 

For a given Z' mass and couplings 
the deviations from the SM can then be calculated and compared with data; since no obvious deviations from the SM 
were observed, LEPII{\cite {LEPEWWG}} places $95\%$ CL lower bounds on Z' masses of $673(481,434,804,1787)$ GeV for the 
$\chi(\psi,\eta$,LRM$(\kappa=1)$,SSM) models assuming $\lambda=1$. Note that since we are far away from the Z' pole 
these results are not sensitive to any particular assumed values for the Z' width as long as it is not too large. 

The process $e^+e^- \to W^+W^-$ can also be sensitive to the existence of a Z', in particular, in the case where there is some 
substantial Z-Z' mixing{\cite {crew}}. The main reason for this is the well-known gauge cancellations among the SM amplitudes 
that maintains unitarity for 
this process as the center of mass energy increases. The introduction of a Z' with Z-Z' mixing induces tiny shifts in the 
W couplings that modifies these cancellations to some extent and unitarity is not completely restored until energies beyond the 
Z' mass are exceeded. As shown by the first authors in Ref.{\cite {crew}}, the leading effects from Z-Z' mixing can be expressed 
in terms of two $s-$dependent anomalous couplings for the $WW\gamma$ and $WWZ$ vertices, \ie, $g_{WW\gamma}=e(1+\delta_\gamma)$ 
and $g_{WWZ}=e(\cot \theta_W+\delta_Z)$ and inserting them into the SM amplitude expressions. The parameters $\delta_{\gamma,Z}$ 
are sensitive to the Z' mass, its leptonic couplings, as well as the Z-Z' mixing angle. In principle, the constraints on 
anomalous couplings from precision measurements can be used to bound the Z' parameters in a model dependent way. However, 
the current data from LEPII{\cite {LEPEWWG}} is not precise enough to get meaningful bounds. More precise data will, of course, be 
obtained at both the LHC and ILC. 

The measurement of the W mass itself can also provides a constraint on $\delta \rho$ since the predicted W mass is altered 
by the fact that $M_Z\neq M_{Z_1}$. Some algebra shows that the resulting mass shift is expected to be  
$\delta M_W =57.6{{\delta \rho}\over {10^{-3}}}$ MeV. Given that $M_W$ is within $\simeq 30$ MeV of the 
predicted SM value and the current size of theory uncertainties{\cite {radcorr}}, strongly suggests that 
$\delta \rho \leq a~few ~10^{-3}$ assuming a light Higgs. This is evidence of small $r$ and/or $\beta$ if a Z' 
is actually present.

Below the Z pole many low energy experiments are sensitive to a Z'. Here we give only two examples: ($i$) The E-158 
Polarized Moller scattering experiment{\cite {Anthony:2005pm}} essentially 
measures $A_{LR}$ which is proportional to a coupling 
combination $\sim -1/2+2x_{eff}$ where $x_{eff}=x_W$+`new physics'. Here $x_W$ is the running value of 
$\sin^2 \theta_W$ at low $Q^2$ which is reliable calculable. For a Z' (assuming no mixing) the `new physics' piece 
is just ${{-1}\over {\sqrt 2 G_F}}{{g_{Z'}^2}\over {M_Z'}^2}v_e'a_e'$, which can be determined in your favorite 
model. Given the data{\cite {Anthony:2005pm}, $x_{eff}-x_W=0.0016\pm 0.0014$, one finds, \eg, that 
$M_{Z_\chi} \geq 960\lambda$ GeV at $90\%$ CL. 
($ii$) Atomic Parity Violation(APV) in heavy atoms measures the effective parity violating interaction between 
electrons and the nucleus and is parameterized via the `weak charge', $Q_W$, which is again calculable in your 
favorite model: 
\begin{equation}
Q_W=-4\sum_i {{M_Z^2}\over {M_{Z_i}^2}}a_{e_i}[v_{u_i}(2Z+N)+v_{d_i}(2N+Z)]\,, 
\end{equation}
$=-N+Z(1-4x_W)+$ a Z' piece,
in the limit of no mixing; here the sum extends over all neutral gauge bosons. 
The possible shift, $\Delta Q_W$, from the SM prediction then constrains Z' parameters. The highest 
precision measurements from $Cs^{133}$ yield{\cite{ Ginges:2003qt} $\Delta Q_W=0.45\pm 0.48$ which then 
imply (at $95\%$ CL) $M_{Z_\chi}>1.05\lambda$ TeV and $M_{Z_{LRM}}>0.67$ TeV for $\kappa=1$. Note that though both these 
measurements take place at very low energies, their relative cleanliness and high precision allows us to probe TeV 
scale Z' masses. Fig.~\ref{fig2} shows the predicted value of the running $\sin^2 \theta_W${\cite {bill}} together with the 
experimental results obtained from E-158, APV and NuTeV{\cite {Zeller:2001hh}}. The apparent 
$\sim 3\sigma$ deviation in the NuTeV result 
remains controversial but is at the moment usually ascribed to our 
lack of detailed knowledge of, \eg, the strange quark parton densities and not to new physics.   
\begin{figure}
\centerline{\psfig{file=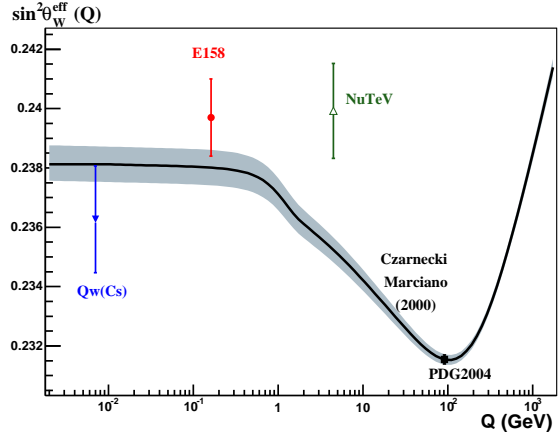,width=8.5cm}}
\caption{A comparison by E-158 of the predictions for the running value of $\sin^2 \theta_W$ with the results of 
several experiments as discussed in the text.}
\label{fig2}
\end{figure}

\subsection{Direct Z' Searches}

In this case we rely on the Drell-Yan process at the Tevatron as mentioned above. The present lack of any signal 
with an integrated luminosity approaching 
$\sim 1$ fb$^{-1}$ allows one to place a model-dependent lower bound on the mass 
of any Z'.  The process $p\bar p \to l^+l^-+X$ at leading order arises from the parton-level subprocess $q\bar q 
\to l^+l^-$ which is quite similar to the $e^+e^- \to f\bar f$ reaction discussed above. The cross section for the 
inclusive process is described by 4 variables: the collider CM energy, $\sqrt s$, the invariant mass of the lepton 
pair, $M$, the scattering angle between the $q$ and the $l^-$, $\theta^*$, and the lepton rapidity in the lab 
frame, $y$, which depends on its energy($E$) and longitudinal momentum($p_z$): $y={1\over {2}}\log 
\Big[{{E+p_z}\over {E-p_z}}\Big]$. For a massless particle, this is the same as the pseudo-rapidity, $\eta$. 
With these variables the triple differential cross section for the Drell-Yan 
process is given by ($z=\cos \theta^*$)     
\begin{equation}
{{d\sigma}\over {dM~dy~dz}}={{K(M)}\over {48\pi M^3}}\sum_q\Big[S_qG_q^+(1+z^2)+2A_qG_q^-z\Big]\,,
\end{equation}
where $K$ is a numerical factor that accounts for NLO and NNLO QCD corrections{\cite {nnlo}} as well as leading electroweak 
corrections{\cite {electroweak} and is roughly of order $\simeq 1.3$ for suitably defined couplings,  
\begin{equation}
G_q^\pm=x_ax_b\Big[q(x_a,M^2)\bar q(x_b,M^2)\pm q(x_b,M^2)\bar q(x_a,M^2)]\,,
\end{equation}
are products of the appropriate parton distribution functions(PDFs), with $x_{a,b}=Me^{\pm y}/\sqrt s$ being the relevant 
momentum fractions, which are evaluated at the scale $M^2$ and 
\begin{eqnarray}
S_q &=& \sum_{ij}P_{ij}(s\to M^2)B_{ij}(f\to q)\\ \nonumber
A_q &=& \sum_{ij}P_{ij}(s\to M^2)C_{ij}(f\to q)\,,
\end{eqnarray}
with $B,C$ and $P$ as given above. In order to get precise limits (and to measure Z' properties once discovered 
as we will see later), the NNLO QCD corrections play an important role{\cite {nnlo}} as do the leading order 
electroweak radiative 
corrections{\cite {electroweak}}. Apart from the machine luminosity errors the largest 
uncertainty in the above cross section is due to the PDFs. For $M\lsim 1$ TeV or so 
these errors are of order $\simeq 5\%${\cite {Houston}} but grow somewhat bigger for 
larger invariant masses: $\sim 15(25)\%${\cite {Belotelov}} for $M=3(5)$ TeV. As a point of comparison 
the corrected SM predictions for the W and Z production cross sections at the Tevatron are seen to 
agree with the data from both CDF and D0 at the level a few percent{\cite {nnlo2}}.

It is somewhat more useful to perform some of the integrals above in order to make direct comparison with experimental 
data. To this end we define (for our LHC discussion below) 
\begin{equation}
{{d\sigma^\pm}\over {dM~dy}} =\Big[\int_0^{z_0} \pm \int_{-z_0}^0\Big]{{d\sigma}\over {dM~dy~dz}}\,,
\end{equation}
and subsequently 
\begin{equation}
{{d\sigma^\pm}\over {dM}} =\Big[\int_{y_{min}}^Y \pm \int_{-Y}^{-y_{min}}\Big]{{d\sigma^\pm}\over {dM~dy}}\,. 
\end{equation}
Here $Y$ is cut representing the edge of the central detector acceptance($\simeq 1.1$ 
for the Tevatron detectors and $\simeq 2.5$ for those at the LHC) 
with $z_0=min[\tanh(Y-|y|),1]$ being the corresponding angular cut. $y_{min}$ is a possible cut employed to define 
the Z' boost direction which we will return to below. As in the case of $e^+e^-$ collisions above, one can define an 
$A_{FB}(M)=d\sigma^-/d\sigma^+$.

\begin{figure}
\centerline{\psfig{file=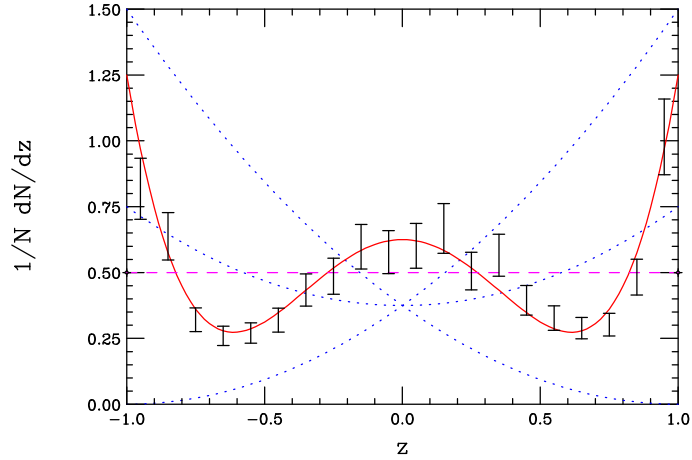,width=6.0cm,angle=90}}
\caption{Normalized leptonic angular distribution predicted from the decay of particles with different spin produced 
in $q\bar q$ annihilation. The dashed(solid,dotted) curves are for spin-0(2,1). The generated data corresponds to 1000 
events in the spin-2 case.}
\label{fig3}
\end{figure}

A Z', being a weakly interacting beast, generally has a rather narrow width to mass ratio, \ie, 
$\Gamma_{Z'}^2/M_{Z'}^2 <<1$; \eg, in the case of the 
SM Z this ratio is $\simeq 10^{-3}$. This being the case, almost the entire Z'  
event rate comes from a rather narrow window of $M$ values: $M\simeq M_{Z'}\pm 2\Gamma_{Z'}$, or so. In this limit 
we can approximate the resonance as a $\delta$-function in $M$ and drop all of the SM contributions to the sums 
above. In this case, pieces of the $P_{ij}$ that go as, \eg, $M^4/|(M^2-M_{Z'}^2)+iM_{Z'}\Gamma_{Z'}|^2$ can be 
replaced by ${\pi \over {2}}\delta(M-M_{Z'}){{M_{Z'}^2}\over {\Gamma_{Z'}}}$, up to $\Gamma_{Z'}^2/M_{Z'}^2$ 
corrections, so that integrals over $M$ can be performed analytically (since the integral over the PDFs is now just 
a constant factor). In such a limit, the contribution to the 
cross section for $l^+l^-$ production from the Z' is just $\sigma_{Z'} B(Z'\to l^+l^-)$ with $\sigma_{Z'}$ being the 
integrated value of the cross section at $M=M_{Z'}$, \ie, at the Z'peak, and $B$ being the leptonic branching fraction of 
the $Z'$. This is called the Narrow Width Approximation(NWA). In a similar way, $A_{FB}$ on the Z' pole in the NWA 
is just the ratio $d\sigma^-/d\sigma^+$ evaluated at $M_{Z'}$; note that this ratio does {\it not} depend upon what 
decay modes (other than leptonic) that the Z' might have. Also note that in the NWA, the continuum 
Drell-Yan background makes no contribution to the event rate. This is a drawback of the NWA since it is 
sometimes important to know the height of the Z'peak relative to this continuum to ascertain the Z' signal significance.  
\begin{figure}[ht]
\begin{center}
  \minifig{2.2in}{\epsfig{figure=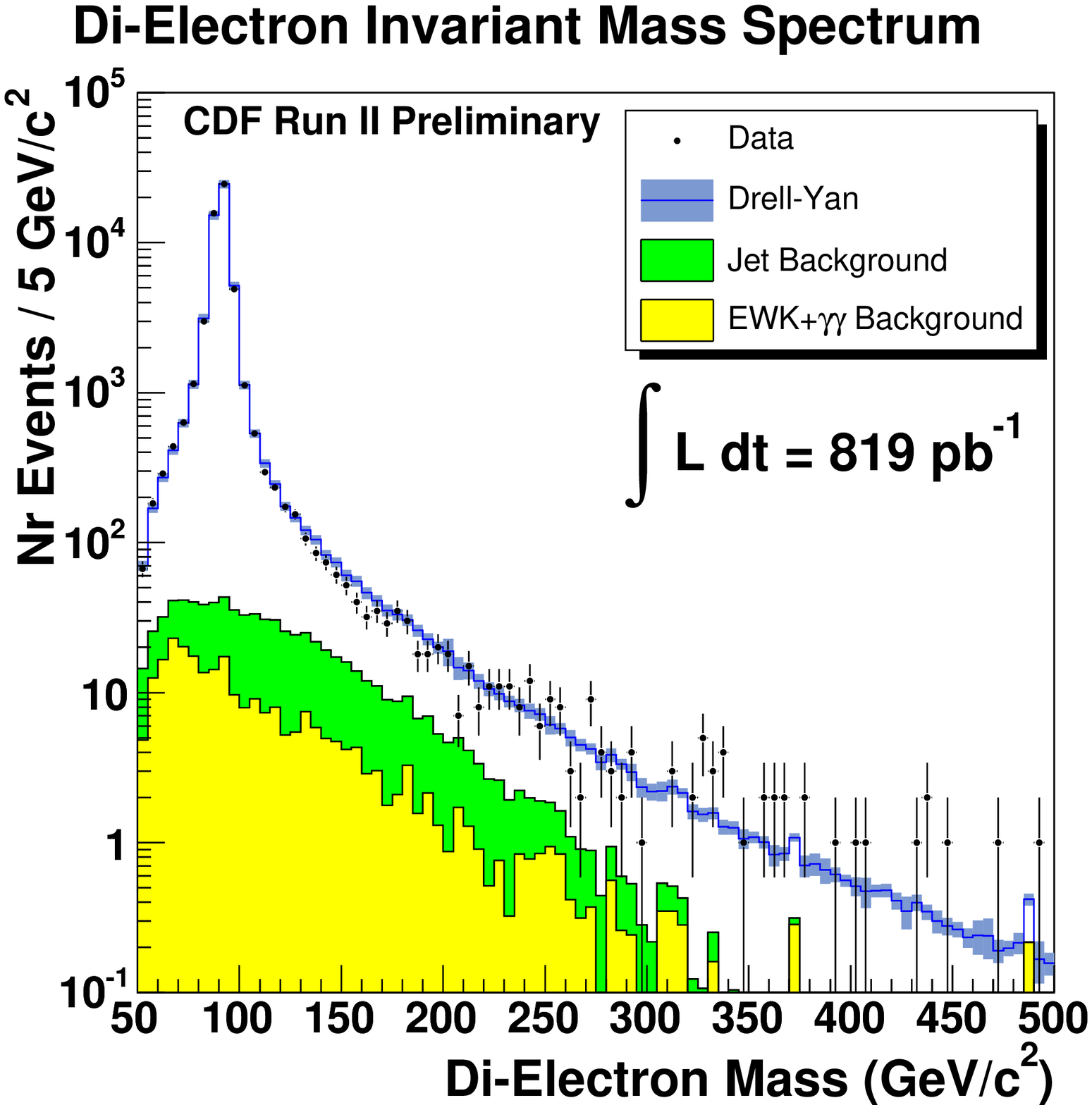,width=2.3in}}{a}
  \hspace*{4pt}
  \minifig{2.1in}{\epsfig{figure=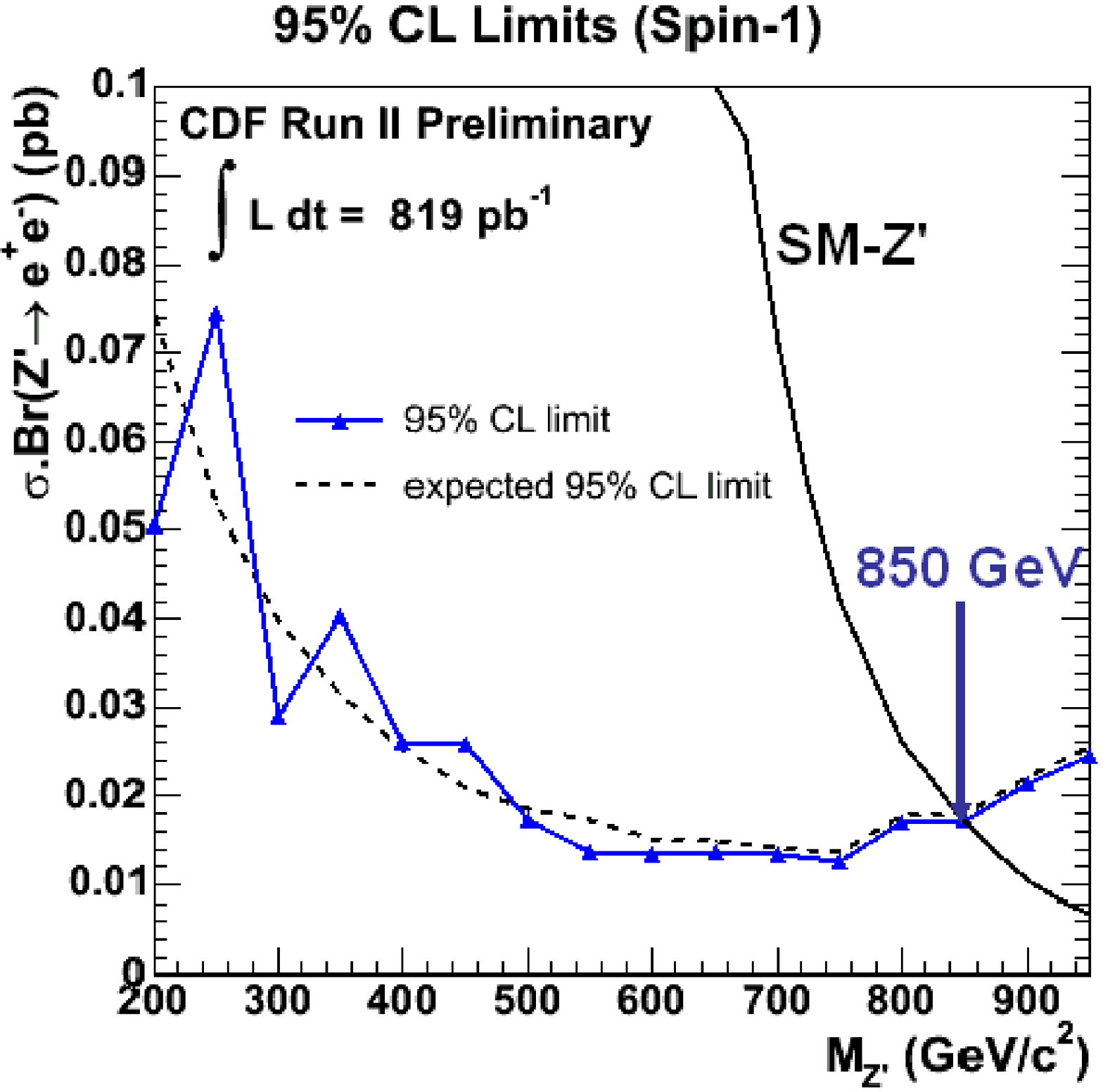,width=2.3in}}{b}
  \caption{(a) The Drell-Yan distribution as seen by CDF. (b) CDF cross section lower bound in comparison to the 
predictions for the Z' in the SSM.}%
  \label{fig4}%
\end{center}
\end{figure}

It is evident from the above cross section expressions that the Z' (as well as $\gamma$ and $Z$) induced Drell-Yan 
cross section involves only terms with a particular angular dependence due to the spin-1 nature of the exchanged particles. 
In the NWA on the Z' pole itself the leptonic angular distribution is seen to behave  
as $\sim 1+z^2 +8A_{FB}z/3$, which is typical of a spin-1 particle. If the Z' had not been a Z' but, say, a $\tilde \nu$ 
in an R-parity violating SUSY model{\cite {slepton}} which is spin-0, then the angular 
distribution on the peak would have been $z$-independent, \ie, 
flat(with, of course, $A_{FB}=0$). This is quite different than the ordinary Z' case. If the Z' had instead been an 
RS graviton{\cite {Randall:1999ee}} with spin-2, then the $q\bar q \to l^+l^-$ part of the cross section would behave as 
$ \sim 1-3z^2+4z^4$, while the $gg \to l^+l^-$ part would go as $\sim 1-z^4$, both parts also yielding $A_{FB}=0$. These 
distributions are also quite distinctive. Fig.~\ref{fig3} shows an example of these (normalized) distributions 
and demonstrates that with less than a few hundred events they are very easily distinguishable. Thus 
the Z' spin should be well established without much of any ambiguity given sufficient luminosity. 
\begin{figure}[ht]
\begin{center}
  \minifig{2.2in}{\epsfig{figure=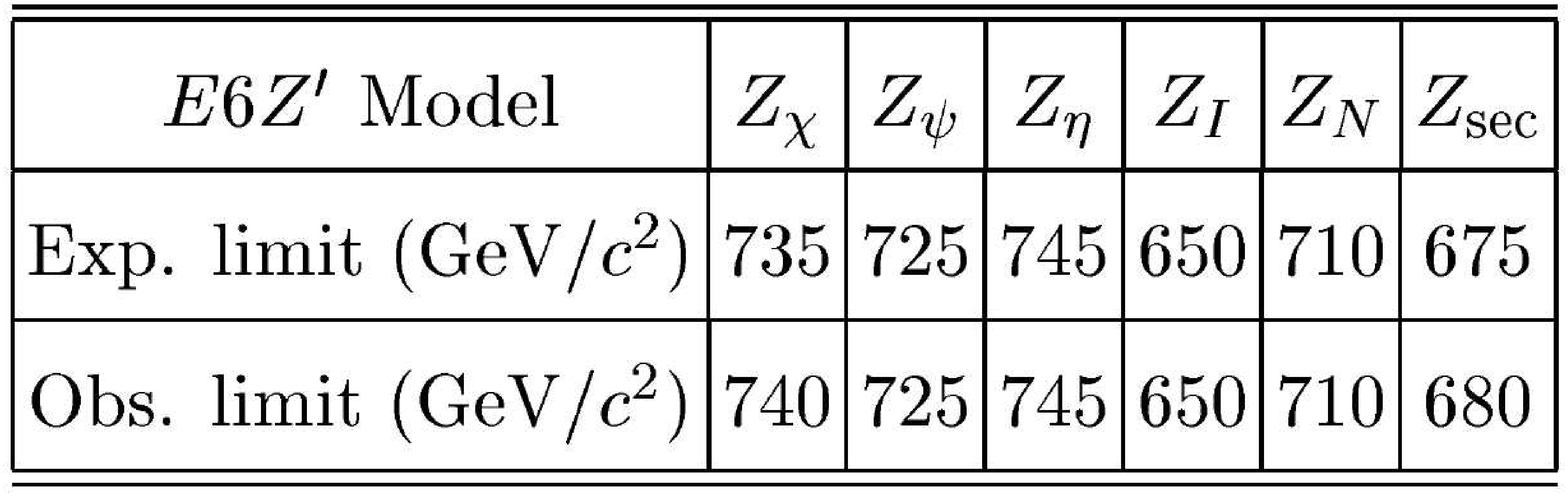,width=2.3in}}{a}
  \hspace*{4pt}
  \minifig{2.1in}{\epsfig{figure=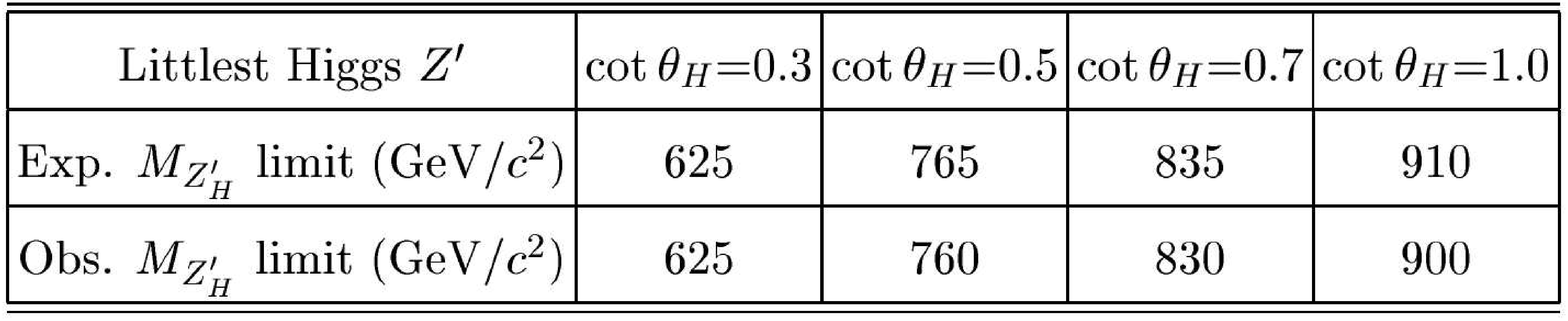,width=2.3in}}{b}
  \caption{Experimental lower bounds from CDF on a number of Z' models: (a) $E_6$ models (b) Little Higgs models.}%
  \label{fig5}%
\end{center}
\end{figure}

An important lesson from the NWA is that the signal rate for a Z' depends upon $B$, the Z' leptonic branching fraction. 
Usually in calculating $B$ one 
assumes that the Z' decays only to SM fields. Given the possible existence of SUSY as well as the additional fermions 
needed in extended electroweak models to cancel anomalies 
this assumption may be wrong. Clearly Z' decays to these other states would 
decrease the value of $B$ making the Z' more difficult to observe experimentally. 

At the Tevatron only 
lower bounds on the mass of a Z' exist. These bounds are obtained by determining the $95\%$ CL upper bound on the 
production cross section for lepton pairs that can arise from new physics as a function of $M(=M_{Z'})$. (Note that 
this has a slight dependence on the assumption that we are looking for a Z' due to the finite acceptance of the 
detector.) Then, for 
any given Z' model one can calculate $\sigma_{Z'} B(Z'\to l^+l^-)$ as a function of $M_{Z'}$ and see at what value of 
$M_{Z'}$ the two curves cross. At present the best limit comes from CDF although comparable limits are also 
obtained by D0{\cite {tevreach}}. The left panel in Fig.~\ref{fig4} shows the 
latest (summer 2006) Drell-Yan spectrum from CDF; the right panel shows the corresponding cross section upper bound 
and the falling prediction for the Z' cross section in the SSM. Here we see that the lower bound is found to 
be 850 GeV {\it assuming} that only SM fermions participate in the Z' decay. For other models an analogous set of theory 
curves can be drawn and the associated limits obtained. 

Fig.~\ref{fig5} shows the resulting constraints (from a different CDF analysis{\cite {Abulencia:2006iv}} 
with a lower integrated 
luminosity but also employing the $A_{FB}$ observable above the mass of the SM Z) on a number of the models 
discussed above all assuming Z' decays to SM particles only and no Z-Z' mixing. Looking at these results we see that 
the Tevatron bounds are generally superior to those from LEPII and are approaching the best that the other precision 
measurements can do. These bounds would degrade somewhat if we allowed the Z' to have additional decay modes; for 
example, if $B$ were reduced by a factor of 2 then the resulting 
search reach would be reduced by 50-100 GeV depending on the model. 

\begin{figure}
\centerline{\psfig{file=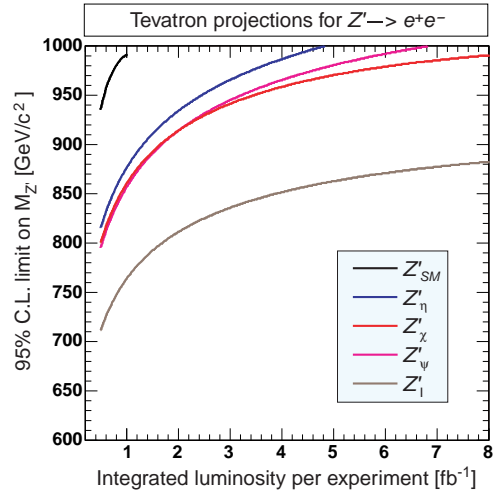,width=6.5cm}}
\caption{Extrapolation of the Z' reach for a number of different models at the Tevatron as the integrated luminosity 
increases. Results from CDF and D0 are combined.}
\label{fig6}
\end{figure}

The Tevatron will, of course, be continuing to accumulate luminosity for several more years possibly reaching as 
high as $8~fb^{-1}$ per experiment. Assuming no signal is found this will increase the Z' search reach lower bound somewhat, 
$\sim 20\%$, as is shown in Fig.~\ref{fig6} from {\cite {future}}. 
At this point the search reach at the Tevatron peters out due to the rapidly falling parton 
densities leaving the mass range above $\sim 1$ TeV for the LHC to explore.

\section{The LHC: Z' Discovery and Identification}

The search for a Z' at the LHC would proceed in the same manner as at the Tevatron. In fact, since the Z' has such 
a clean (\ie, dilepton) signal and a sizable cross section it could be one of the first new physics signatures 
to be observed at the LHC even at relatively low integrated luminosities{\cite{Alemany,Willocq,Baker}}. Fig.~\ref{fig7} shows 
both the theoretical anticipated $95\%$ CL lower bound and the $5\sigma$ discovery reach for several different Z' models at 
the LHC for a single leptonic 
channel as the integrated luminosity is increased; these results are mirrored in detectors studies{\cite {Cousins1}}. 
Here we see that with only $10-20~pb^{-1}$ the LHC detectors will 
clean up any of the low mass region left by the Tevatron below 1 TeV and may actually discover a 1 TeV Z' with 
luminosities in the $30-100~pb^{-1}$ range! In terms of discovery, however, to get out to the $\sim 4-5$ TeV mass range 
will requite $\sim 100~ fb^{-1}$ of luminosity. At such luminosities, the $95\%$ CL bound exceeds the $5\sigma$ 
discovery reach by about 700 GeV. In these plots, we have again assumed that the Z' leptonic branching fraction 
is determined by decays only to SM fermions. Reducing $B$ by a factor of 2 could reduce these reaches by $\simeq 10\%$ which 
is not a large effect. 
\begin{figure}[ht]
\begin{center}
  \minifig{2.2in}{\epsfig{figure=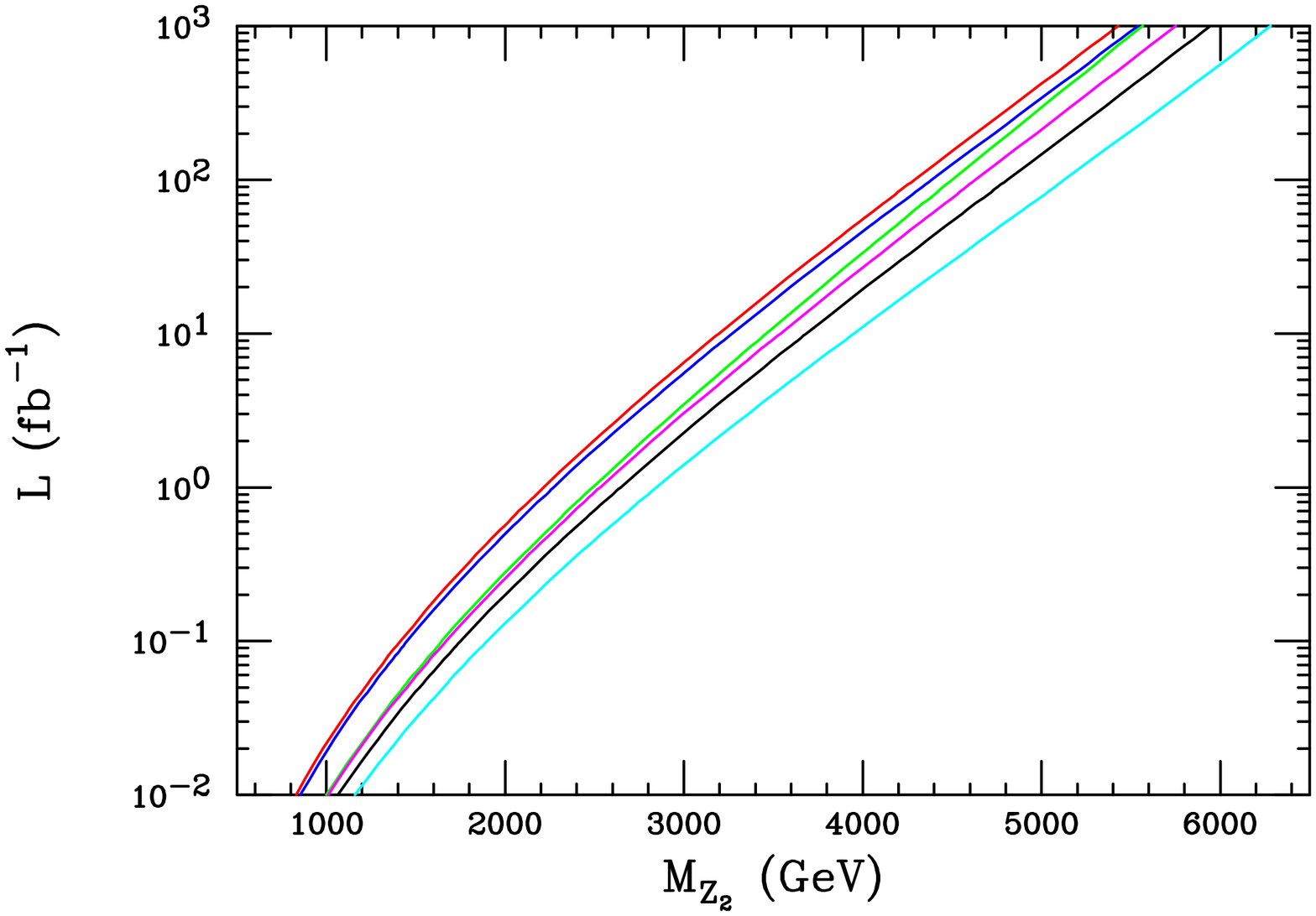,width=1.6in,angle=90}}{a}
  \hspace*{4pt}
  \minifig{2.1in}{\epsfig{figure=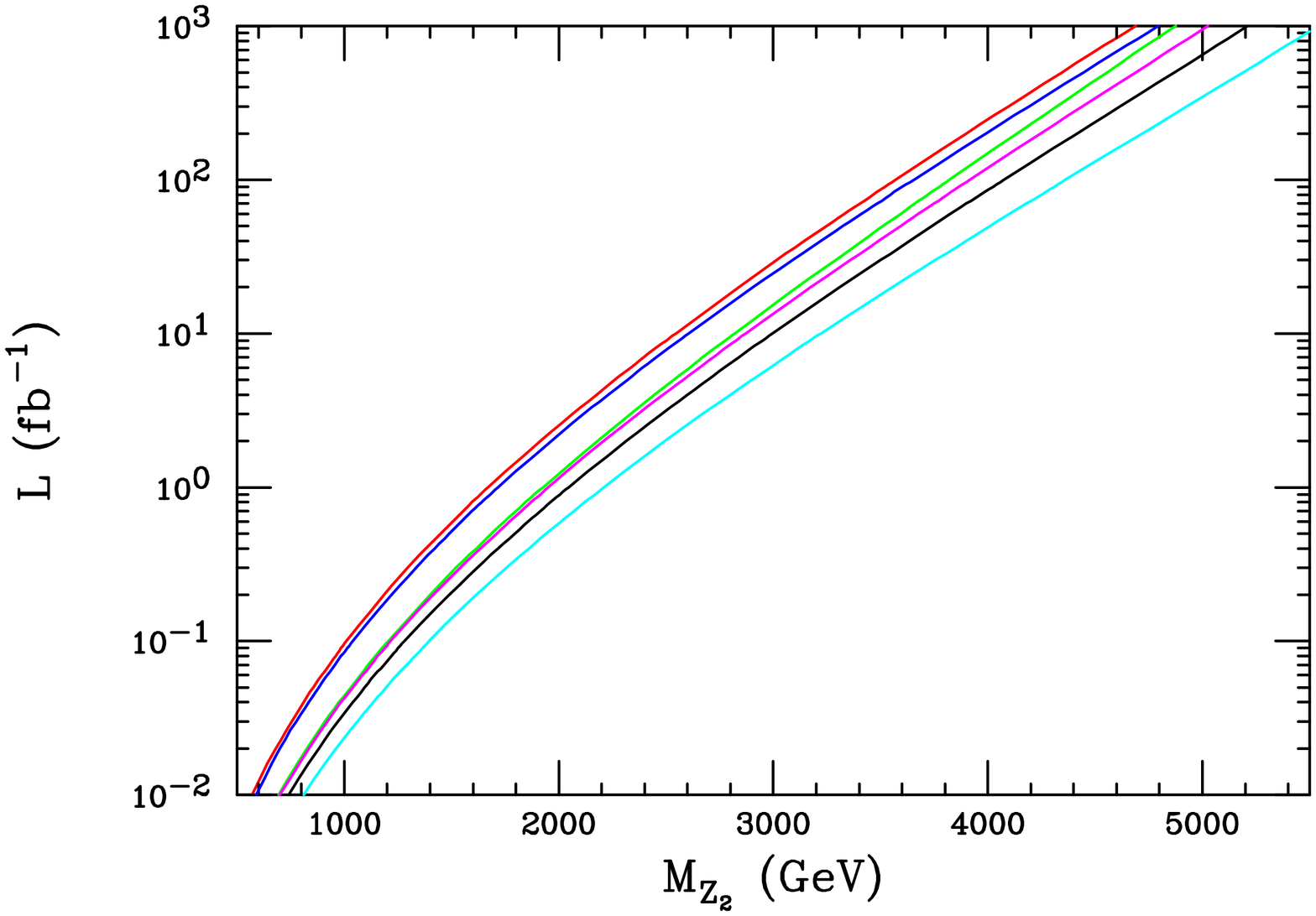,width=1.6in,angle=90}}{b}
  \caption{(a) $95\%$ CL lower bound and (b) $5\sigma$ discovery reach for a Z' as a function of the 
integrated luminosity at the LHC for $\psi$(red), $\chi$(green), $\eta$(blue), the LRM with 
$\kappa=1$(magenta), the SSM(cyan) and the ALRM(black). Decays to only SM fermions is assumed.}%
  \label{fig7}%
\end{center}
\end{figure}

The Z' peak at the LHC should be relatively easy to spot since the SM backgrounds are well understood as 
shown{\cite {Willocq,Azuelos:2004dm}} in Fig.~\ref{fig8}  
for a number of different Z' models. The one problem that may arise is for the case where the Z' 
width, $\Gamma_{Z'}$, is far smaller than the experimental dilepton pair mass resolution, $\delta M$. Typically in most  
models, $\Gamma_{Z'}/M_{Z'}$ is of order $\simeq 0.01$ which is comparable to dilepton pair mass resolution, 
$\delta M/M$, for both ATLAS{\cite{ATLASTDR}} and CMS{\cite {CMSTDR}}. If, however,  
$\Gamma_{Z'}/M_{Z'}<<\delta M/M$, then the Z' resonance is smeared out due to the resolution and the cross section 
peak is reduced by roughly a factor of 
$\sim \Gamma_{Z'}/\delta M$ making the state difficult to observe. This could happen, \eg, if the 
Z' (before mixing with the SM Z) had no couplings to SM fields{\cite {Kumar:2006yg}}.
\begin{figure}[ht]
\begin{center}
  \minifig{2.2in}{\epsfig{figure=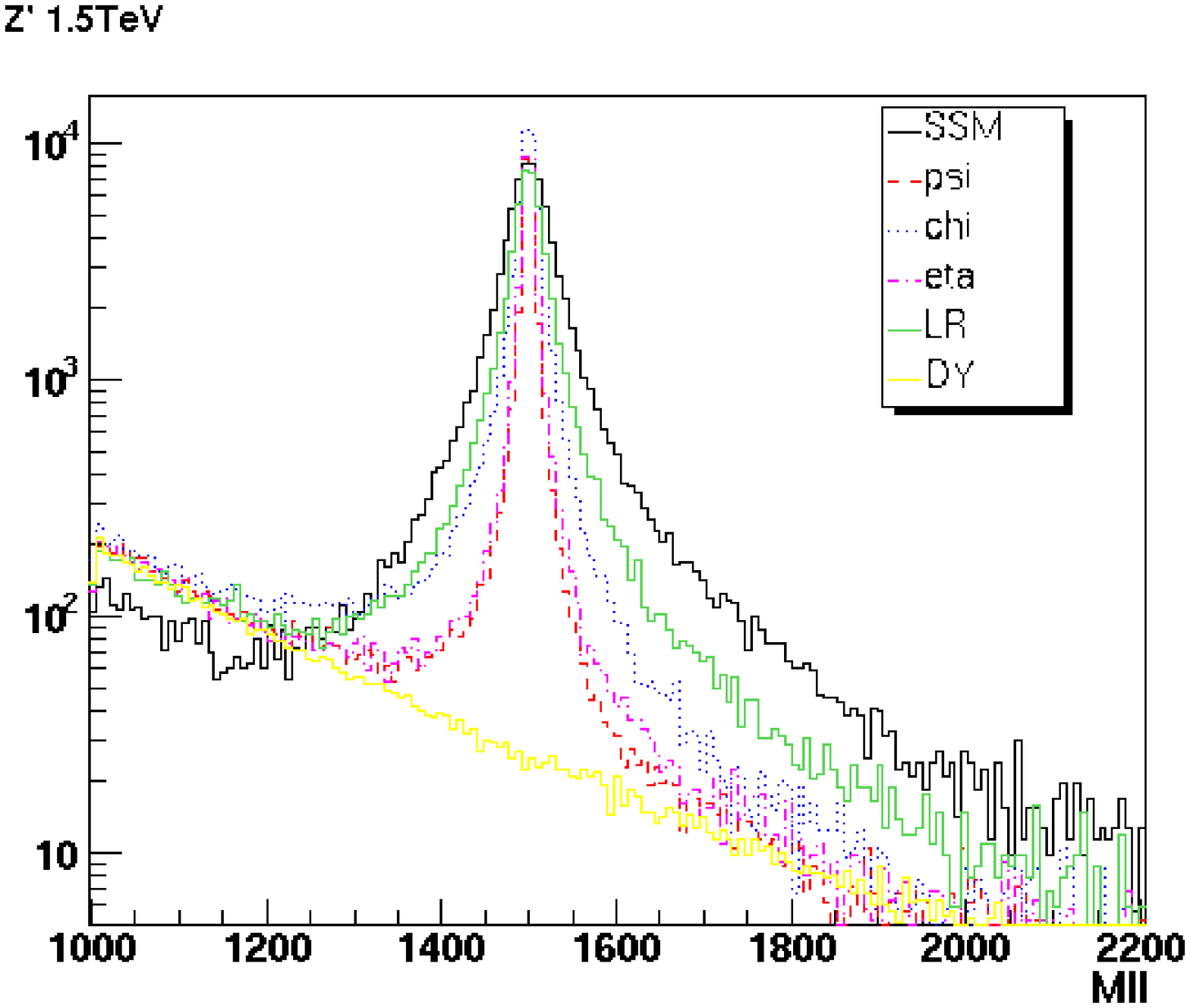,width=2.6in}}{a}
  \hspace*{4pt}
  \minifig{2.1in}{\epsfig{figure=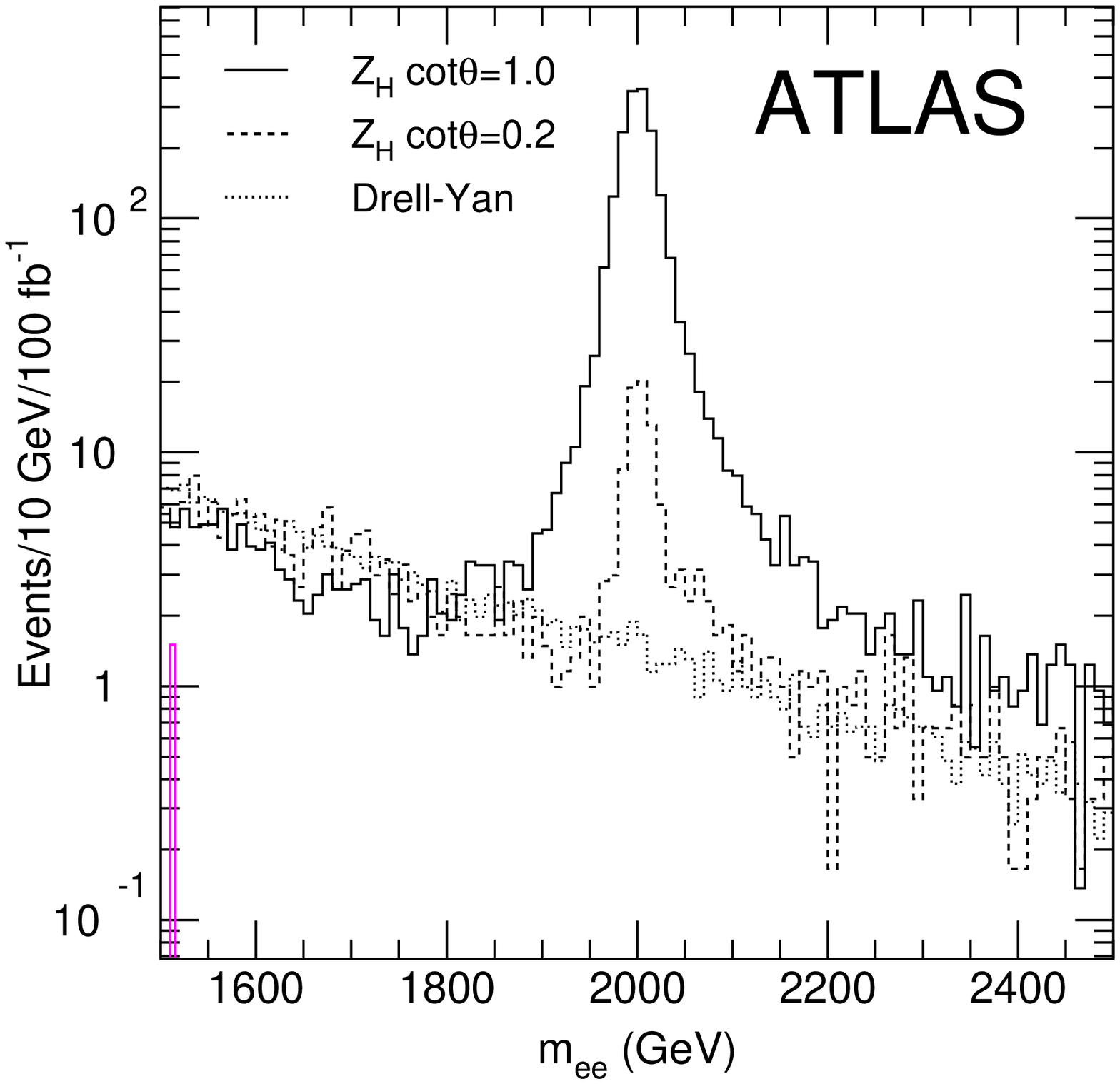,width=2.2in}}{b}
  \caption{Resonance shapes for a number of Z' models as seen by ATLAS assuming $M_{Z'}=1.5$ TeV. The continuum is the 
SM Drell-Yan background.}%
\label{fig8}
\end{center}
\end{figure}

Given the huge mass reach of the LHC it is important to entertain the question of how to `identify' a particular Z' 
model once such a particle is found. This goes beyond just being able to tell the Z' of Model A from the Z' from 
model B. As alluded to in the introduction, if a Z'-like object is discovered, the first 
step will be to determine its spin. Based on the theoretical discussion above this would seem to be rather 
straightforward and studies of this issue have been performed by both ATLAS{\cite {Allanach:2000nr}} 
and CMS{\cite {Cousins:2005pq}}. Generally, 
one finds that discriminating a spin-1 or spin-2 object from one of spin-0 requires several times more events than 
does discriminating spin-2 from spin-1. The requirement of a few hundred events, however, somewhat limits the mass 
range over which such an analysis can be performed. If a particular Z' model has an LHC search reach of 4 TeV, then 
only for masses below $\simeq 2.5-3$ TeV will there be the statistics necessary to perform a reliable spin 
determination. Fig.~\ref{fig9} shows two sample results from this spin analysis. For the ATLAS study in the left 
panel{\cite {Allanach:2000nr}} the lepton angular distribution 
for a weakly coupled 1.5 TeV KK RS graviton is compared with the 
expectation for a SSM Z' of identical mass assuming a luminosity of 100 $fb^{-1}$. Here one clearly sees the obvious 
difference and the spin-2 nature of the resonance. In the right panel{\cite {Cousins:2005pq} the results of a CMS analysis is 
presented with the distinction of a 1.5 TeV Z' and a KK graviton again being considered.  Here one asks for 
the number of events($N$) necessary to distinguish the two cases, at a fixed number of standard deviations, 
$\sigma$, which is seen to 
grow as (as it should ) with $\sqrt N$. For example, a $3\sigma$ separation is seen to require $\simeq 300$ events. 
\begin{figure}[ht]
\begin{center}
  \minifig{2.2in}{\epsfig{figure=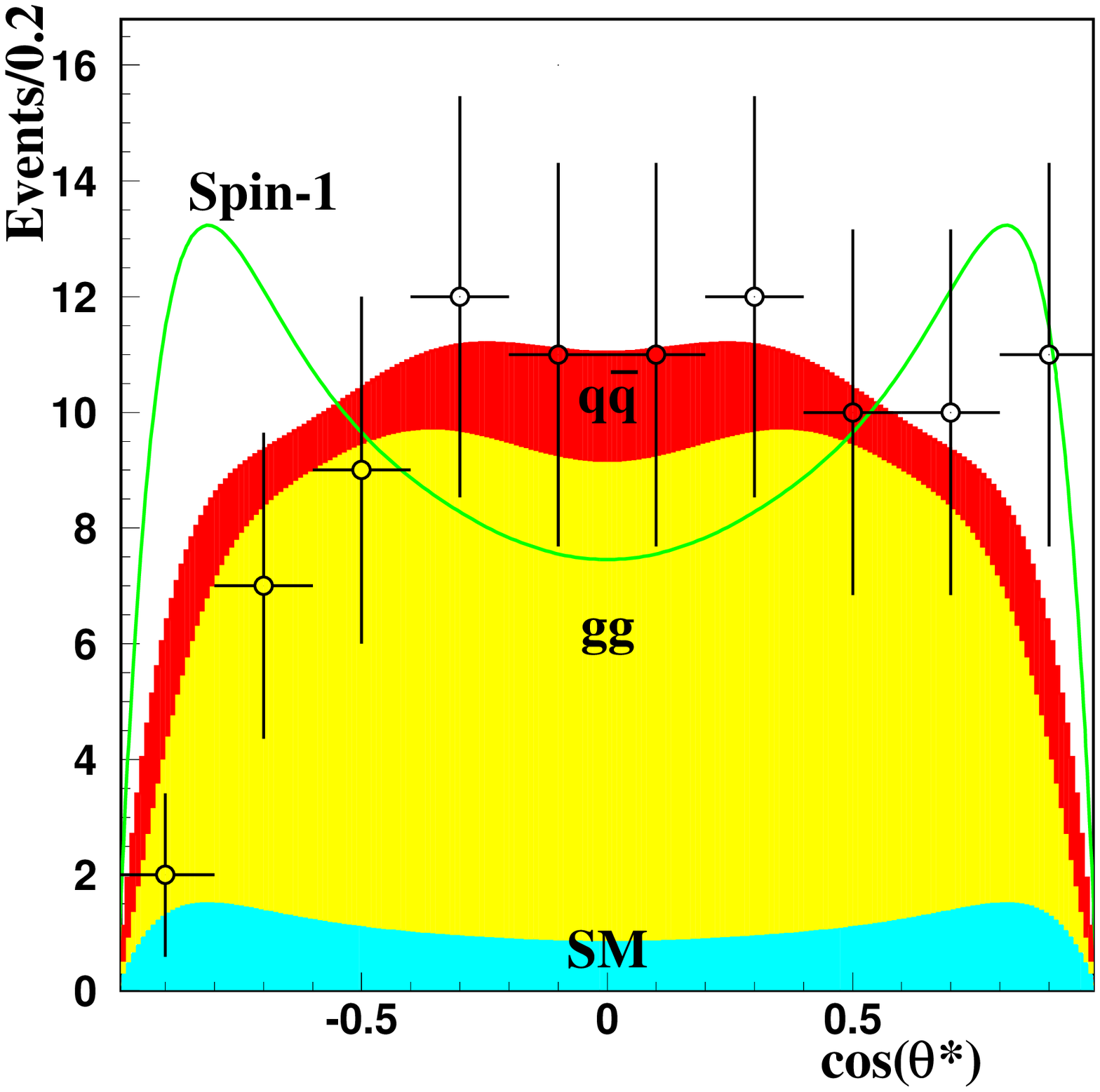,width=2.3in}}{a}
  \hspace*{4pt}
  \minifig{2.1in}{\epsfig{figure=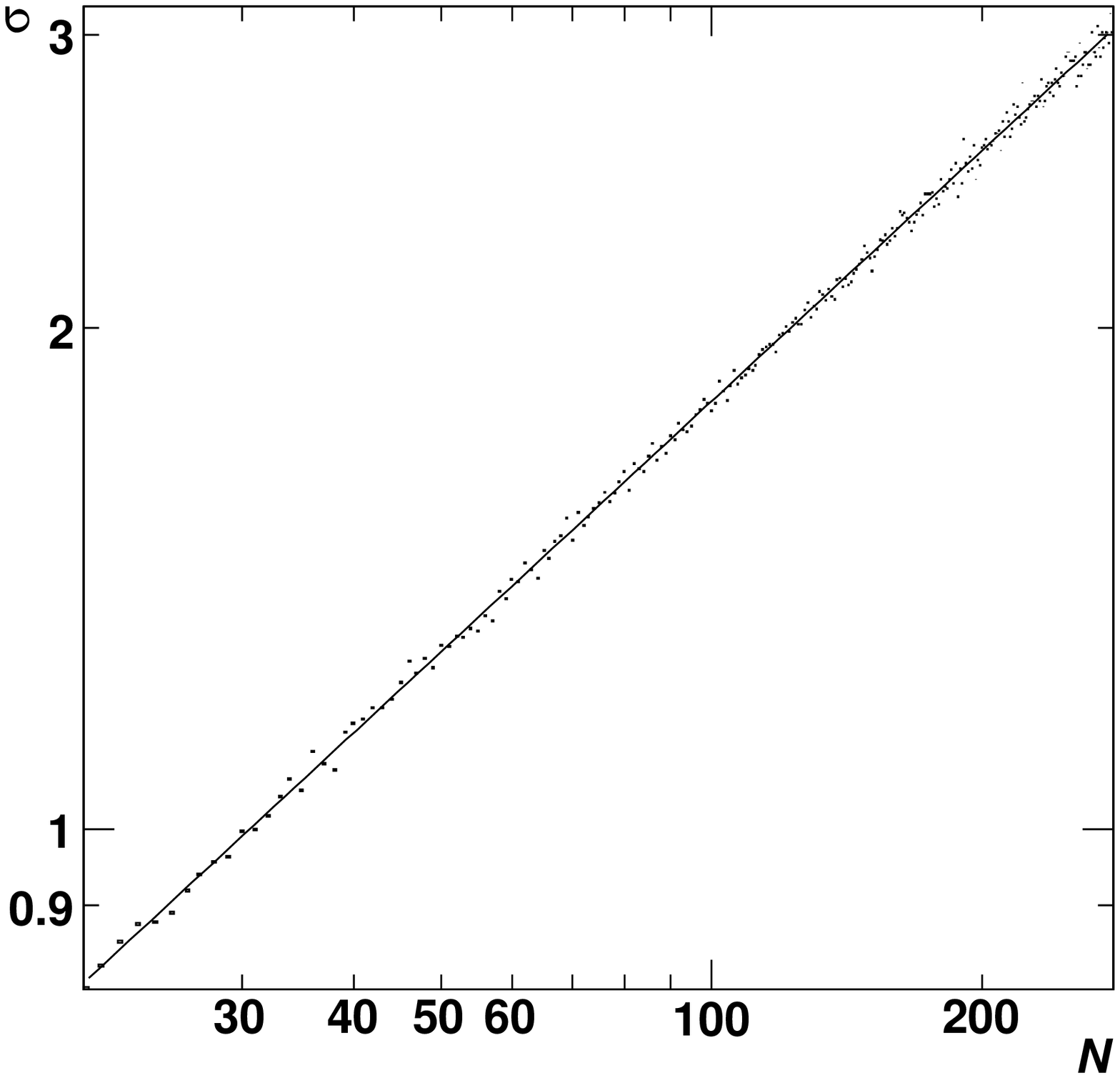,width=2.3in}}{b}
  \caption{(a) The theoretical predictions for 1.5 TeV SSM Z' and RS graviton resonance shapes at ATLAS in comparison to 
the graviton signal data. (b) Differentiation, in $\sigma$, of spin-1 and spin-2 resonances at CMS as a function of the number 
of events assuming a 1.5 TeV mass.}%
  \label{fig9}%
\end{center}
\end{figure}

Once we know that we indeed have a spin-1 object, we next need to `identify' it, \ie, uniquely 
determine its couplings to the various SM fermions. (Note that almost all LHC experimental analyses up to now have 
primarily focused on being able to distinguish models and not on actual coupling extractions.) We would like 
to be able to do this in as model-independent a way as possible, \eg, we should not assume that the Z' decays only 
to SM fields. Clearly this task will require many more events than a simple discovery or even a spin determination and 
will probably be difficult 
for a Z' with a mass much greater than $\simeq 2-2.5$ TeV unless integrated luminosities significantly in excess of 100 
$fb^{-1}$ are achieved (as may occur at the LHC upgrade{\cite {Gianotti:2002xx}}). 
Some of the required information can be obtained using 
the dilepton (\ie, $e^+e^-$ and/or $\mu^+\mu^-$) discovery channel but to obtain more information the examination of 
additional channels will also be necessary. 
\begin{figure}[ht]
\begin{center}
  \minifig{2.2in}{\epsfig{figure=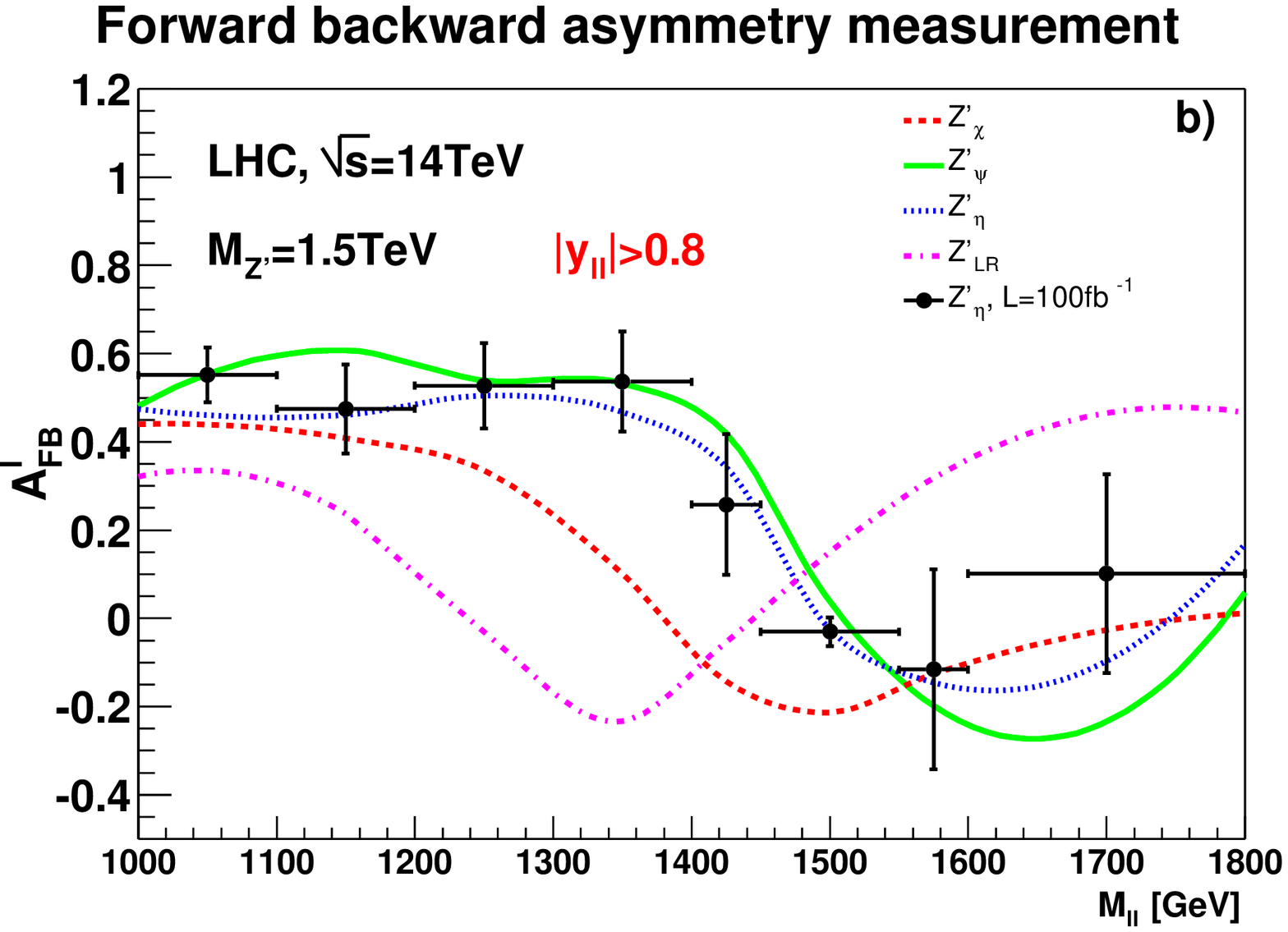,width=2.4in}}{a}
  \hspace*{4pt}
  \minifig{2.1in}{\epsfig{figure=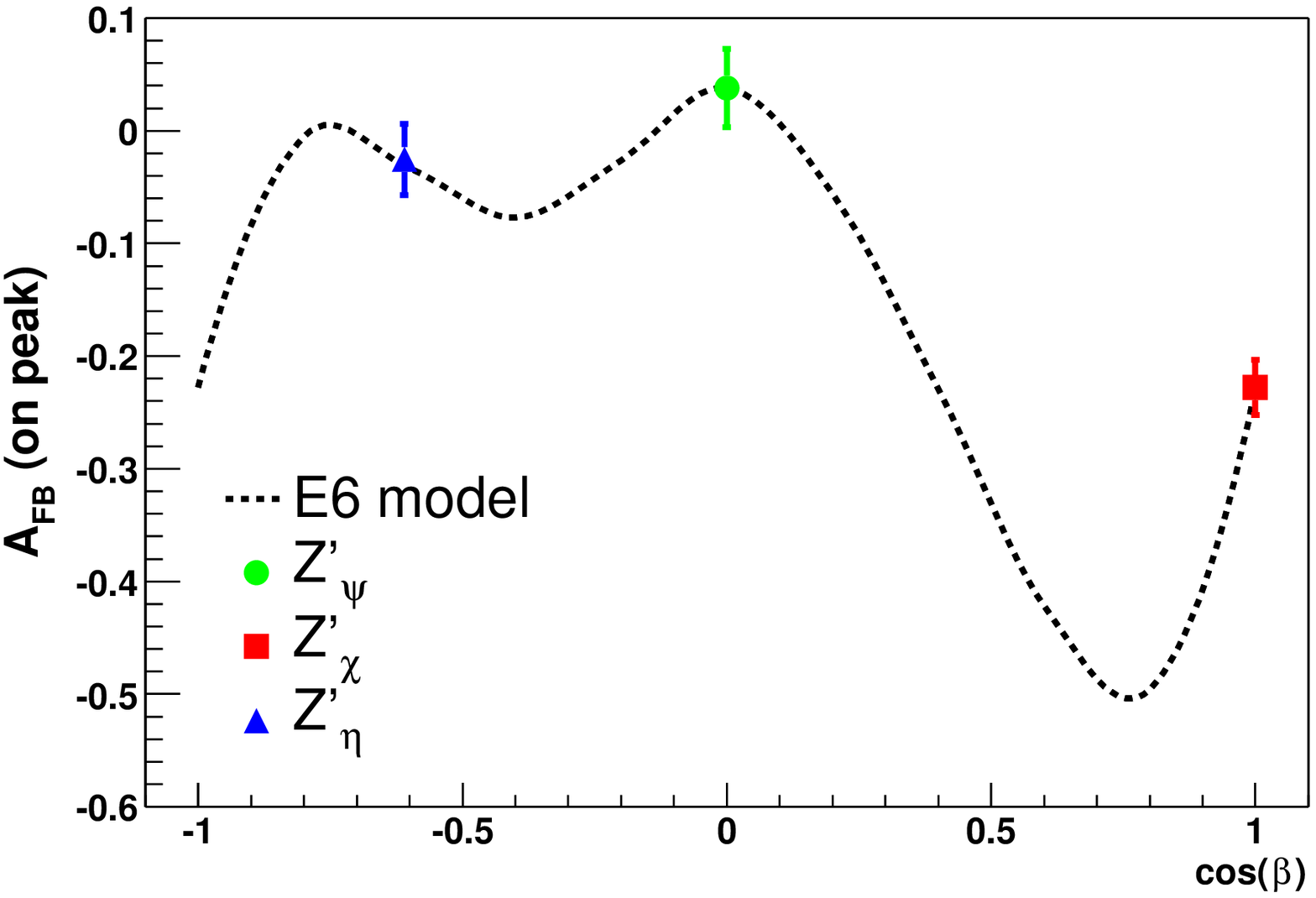,width=2.4in}}{b}
  \caption{(a)$A_{FB}$ near a 1.5 TeV Z' in a number of models. (b)On-peak differentiation of $E_6$ models using 
 $A_{FB}$ showing statistical errors for a 1.5 TeV Z'.}%
  \label{fig10}%
\end{center}
\end{figure}
\begin{table}[ht]
\tbl{Results on $\sigma_{ll}$ and $\sigma_{ll} \times \Gamma_{Z'}$ for all studied models from ATLAS. Here one compares the 
input values from the generator with the reconstructed values obtained after full detector simulation.}
{\begin{tabular}{ccccccc}\hline
 \multicolumn{2}{c}{}  & $\sigma^{gen}_{ll}$(fb) & $\sigma^{rec}_{ll}$(fb)& $\sigma^{rec}_{ll} \times \Gamma_{rec}$
(fb.GeV)\\ \hline
&$SSM$                              & 78.4$\pm$0.8 & 78.5$\pm$1.8 & 3550$\pm$137 \\ \cline{2-5}
&$\psi$                             & 22.6$\pm$0.3 & 22.7$\pm$0.6 & 166$\pm$15 \\ \cline{2-5}
$M=1.5$\,TeV&$\chi$                 & 47.5$\pm$0.6 & 48.4$\pm$1.3 & 800$\pm$47 \\ \cline{2-5}
&$\eta$                             & 26.2$\pm$0.3 & 24.6$\pm$0.6 & 212$\pm$16 \\ \cline{2-5}
&$LR$                               & 50.8$\pm$0.6 & 51.1$\pm$1.3 & 1495$\pm$72 \\ \hline \hline
\raisebox{-1.5ex}{$M=4$\,TeV}&$SSM$ & 0.16$\pm$0.002& 0.16$\pm$0.004& 19$\pm$1 \\ \cline{2-5}
                             &$KK$  & 2.2$\pm$0.07& 2.2$\pm$0.12& 331$\pm$35 \\ \hline
\end{tabular}}
\label{tbl2}
\end{table}

In the dilepton mode, three obvious observables present themselves: ($i$) the cross section, $\sigma_{ll}$, on and 
below the Z' peak (it is generally very small above the peak), ($ii$) the corresponding values of $A_{FB}$ and ($iii$) 
the width, $\Gamma_{Z'}$, of the Z' from resonance peak shape measurements. Recall 
that while $A_{FB}$ is $B$ insensitive, both $\sigma_{ll}$ and  $\Gamma_{Z'}$ {\it are} individually 
sensitive to what we assume about the leptonic branching fraction, $B$, so 
that they cannot be used independently. In the NWA, however, one sees that the product of the peak cross section 
and the Z' width, $\sigma_{ll}\Gamma_{Z'}$, is {\it independent} of $B$. (Due to smearing and finite width effects, one 
really needs to take the product of $d\sigma^+/dM$, integrated around the peak and $\Gamma_{Z'}$.) Table~\ref{tbl2} 
from an ATLAS study {\cite {Schafer}} demonstrates that the product $\sigma_{ll}\Gamma_{Z'}$ can be reliably determined 
at the LHC in full simulation, reproducing well the original input generator value. 
\begin{figure}
\centerline{\psfig{file=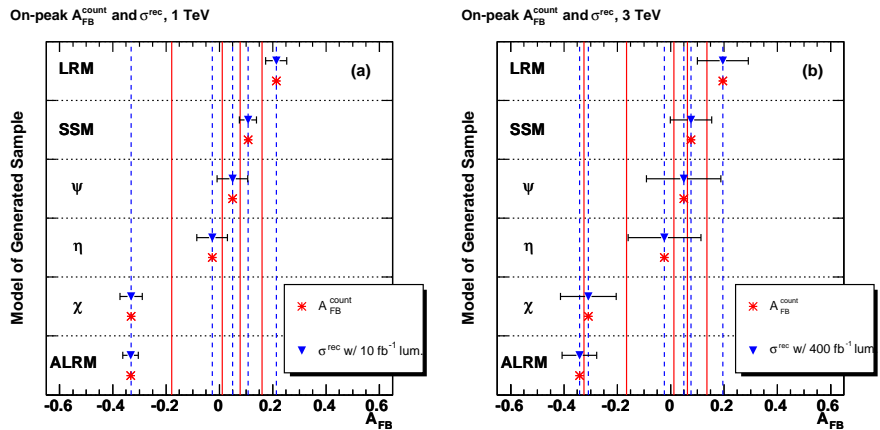,width=12.0cm}}
\caption{CMS analysis of Z' model differentiation employing $A_{FB}$ assuming $M_{Z'}=1$ or 3 TeV.}
\label{fig11}
\end{figure}

Let us now consider the quantity $A_{FB}$. At the theory level, the angle $\theta^*$ employed above is defined to 
be that between 
the incoming $q$ and the outgoing $l^-$. Experimentally, though the lepton can be charge signed with relative ease, 
it is 
not immediately obvious in which direction the initial quark is going, \ie, to determine which proton it came from. 
However, since the $q$ valence distributions are `harder' (\ie, have higher 
average momentum fractions) than the `softer' $\bar q$ sea partons, it is likely{\cite {Haber:1984gd}} that the Z' boost 
direction will be that of the original $q$. Of course, this is not {\it always} true so that making this assumption 
dilutes the true value of $A_{FB}$ as does, \eg, additional gluon radiation. For the Z' to be boosted, 
the leptons in the final state need to have (significant) rapidity, hence the 
lower bound in the integration of the cross section expression above. Clearly, a full analysis needs to take these 
and other experimental issues into account. 

\begin{table}[ht]
\tbl{Measured on-peak $A_{FB}$ for all studied models in the central mass bin from ATLAS. Here the raw value obtained before 
dilution corrections is labeled as `Observed'.}
{\begin{tabular}{|c|c|c|c|c|}\hline
Model & $\int\mathcal{L} (fb^{-1})$ & Generation & Observed & Corrected \\ \hline
1.5\,TeV & \multicolumn{4}{|c|}{} \\ \hline 
$SSM$  & 100 & $+0.088 \pm 0.013 $ & $+0.060 \pm 0.022 $ & $+0.108 \pm 0.027$  \\ \hline
$\chi$ & 100 & $-0.386 \pm 0.013 $ & $-0.144 \pm 0.025 $ & $-0.361 \pm 0.030$  \\ \hline 
$\eta$ & 100 & $-0.112 \pm 0.019 $ & $-0.067 \pm 0.032 $ & $-0.204 \pm 0.039$  \\ \hline 
$\eta$ & 300 & $-0.090 \pm 0.011 $ & $-0.050 \pm 0.018 $ & $-0.120 \pm 0.022$  \\ \hline 
$\psi$ & 100 & $+0.008 \pm 0.020 $ & $-0.056 \pm 0.033 $ & $-0.079 \pm 0.042$  \\ \hline 
$\psi$ & 300 & $+0.010 \pm 0.011 $ & $-0.019 \pm 0.019 $ & $-0.011 \pm 0.024$  \\ \hline
$LR$   & 100 & $+0.177 \pm 0.016 $ & $+0.100 \pm 0.026 $ & $+0.186 \pm 0.032$  \\ \hline 
4\,TeV  & \multicolumn{4}{|c|}{} \\ \hline 
$SSM$  & 10000 & $+0.057 \pm 0.023 $  & $-0.001 \pm 0.040 $ & $+0.078 \pm 0.051$  \\ \hline
$KK$   & 500   & $+0.491 \pm 0.028 $  & $+0.189 \pm 0.057 $ & $+0.457 \pm 0.073$  \\ \hline
\end{tabular}}
\label{tbl3}
\end{table}

The left panel of Fig.~\ref{fig10}  
shows{\cite {Dittmar:2003ir}} $A_{FB}$ as a function of $M$ in the region near a 1.5 TeV Z' for $E_6$ 
model $\eta$ in comparison with the predictions of several other models. Here we see several features, the first being 
that the errors on $A_{FB}$ are rather large except on the Z' pole itself due to relatively low statistics even with 
large integrated luminosities of 100 $fb^{-1}$; this is particularly true above the resonance.  Second, it is clear 
that $A_{FB}$ both on and off the peak does show some reasonable model sensitivity as was hoped. From the right 
panel{\cite {Dittmar:2003ir}} of Fig.~\ref{fig10}} it is clear that the 
various special case models of the $E_6$ family are distinguishable. This 
is confirmed by more detailed studies performed by both ATLAS{\cite {Schafer}} and CMS{\cite {Cousins2}}. 
Fig.~\ref{fig11} from  
CMS{\cite {Cousins2}} shows how measurements of the on-peak $A_{FB}$ can be used to distinguish models with 
reasonable confidence given sufficient statistics (and in the absence of several systematic effects). 
Table~\ref{tbl3} from the ATLAS study{\cite {Schafer}} shows that the original input generator value of the 
on-peak $A_{FB}$ can be reasonably well reproduced with a full detector simulation, taking dilution and other effects 
into account. 
\begin{figure}[ht]
\begin{center}
  \minifig{2.2in}{\epsfig{figure=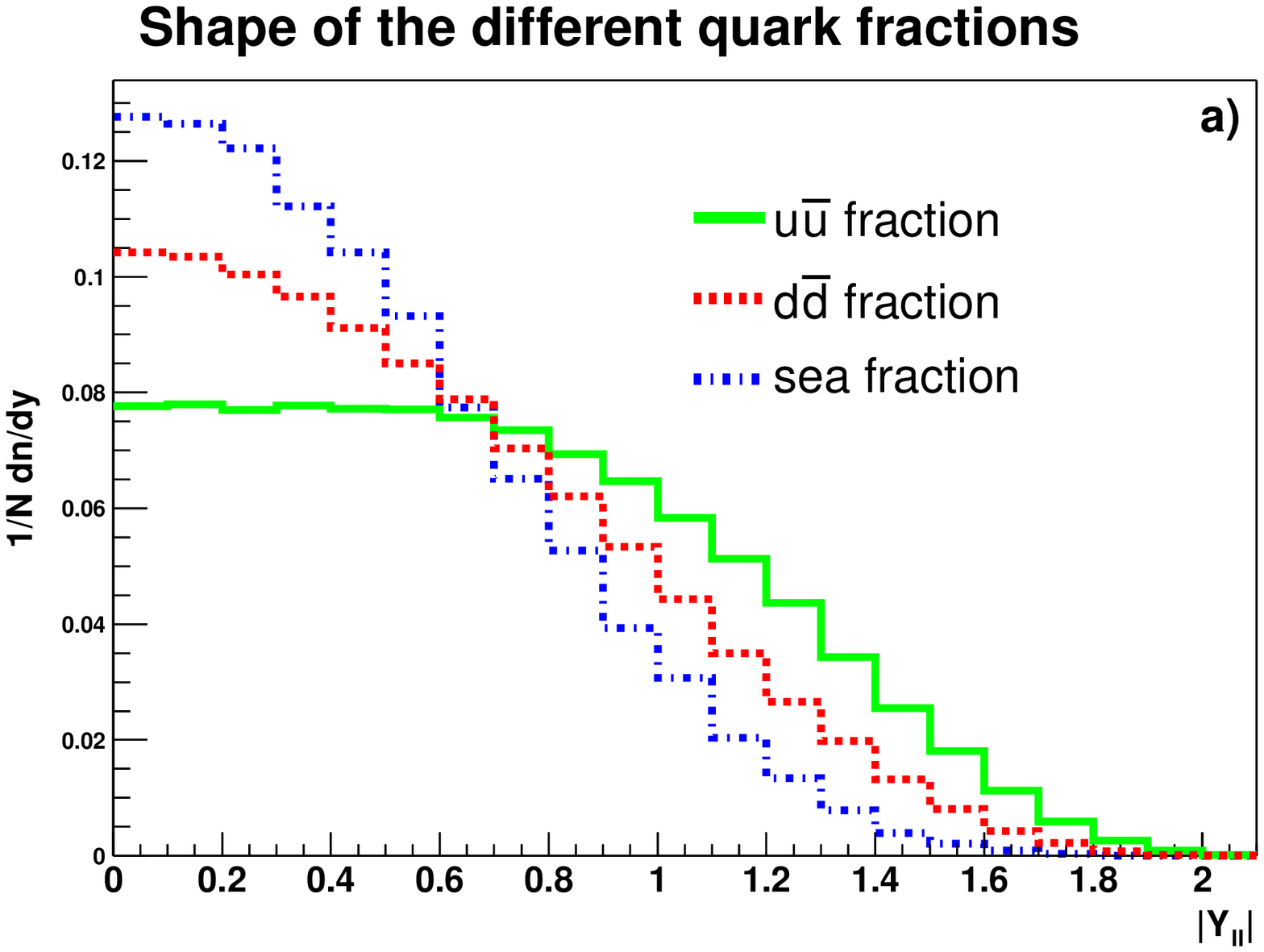,width=2.4in}}{a}
  \hspace*{4pt}
  \minifig{2.1in}{\epsfig{figure=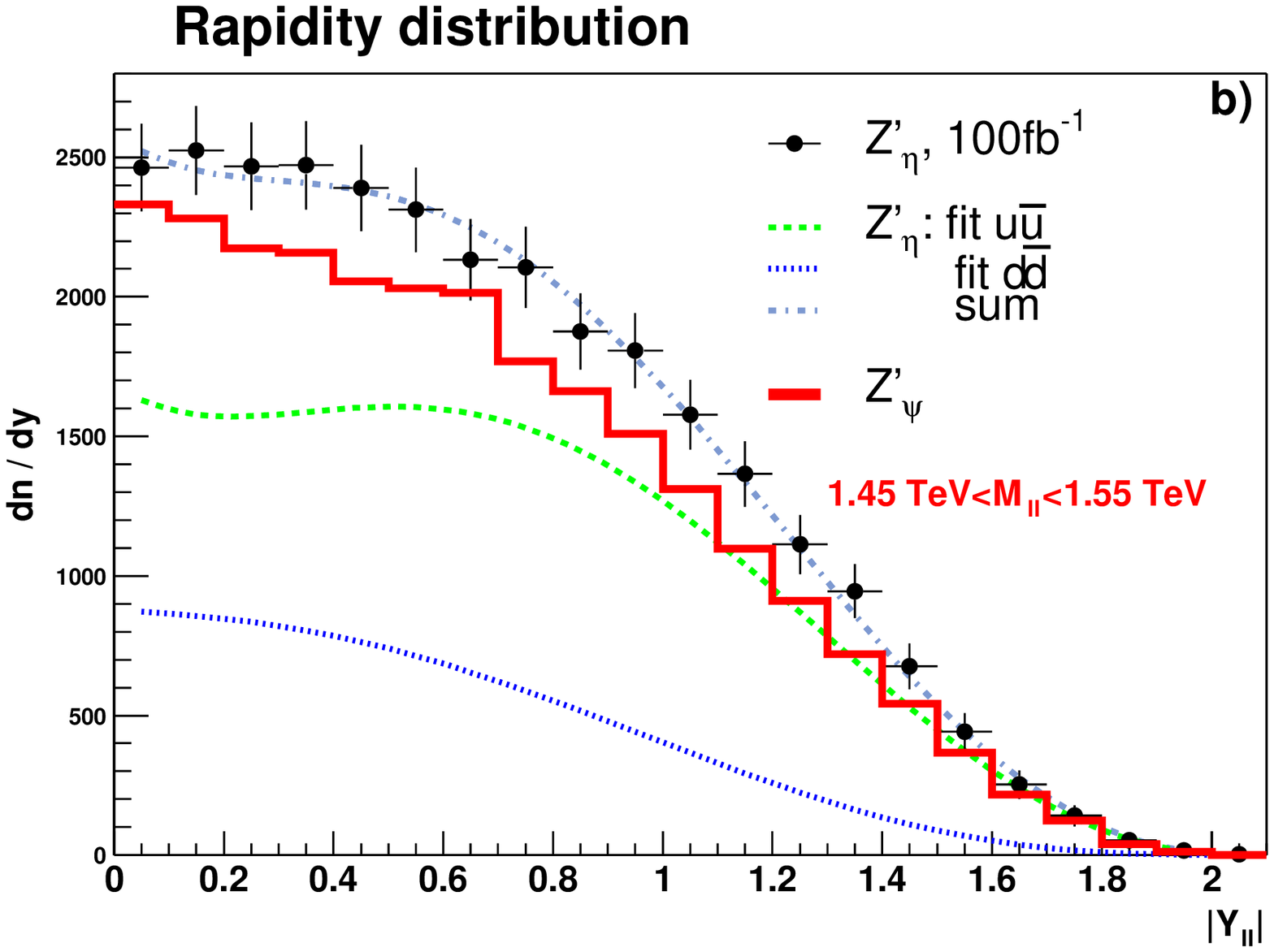,width=2.4in}}{b}
  \caption{(a)Rapidity distributions for different $q\bar q$ induced events. (b)Rapidity distribution differentiation of 
Z' models.}
\label{fig12}
\end{center}
\end{figure}

If a large enough on-peak data sample is available, examining $A_{FB}$ as a function 
of the lepton rapidity{\cite {Rosner:1986cv}}  
can provide additional coupling information. The reason for this is that $u$ and $d$ quarks have different $x$ 
distributions so that the weight of $u\bar u$ and $d\bar d$ induced Z' events changes as the rapidity varies. No 
detector level studies of this have yet been performed. 

Off-peak measurements of $A_{FB}$ are also useful although in this case systematics are more important; as shown in the 
ATLAS study{\cite {Schafer}}, whose results are shown in Table~\ref{tbl4}, it is more difficult to reproduce the 
input generator value of this quantity than in the on-peak case. 
\begin{table}[ht]
\tbl{Measured off peak, $0.8 < M<1.4$ TeV, $A_{FB}$ for all studied models from ATLAS using the same nomenclature as above.}
{\begin{tabular}{|c|c|c|c|c|}\hline
Model & $\int\mathcal{L} (fb^{-1})$ & Generation & Observed & Corrected \\ \hline
1.5\,TeV & \multicolumn{4}{|c|}{} \\ \hline 
$SSM$  & 100 & $+0.077 \pm 0.025 $ & $+0.086 \pm 0.038$ & $+0.171 \pm 0.045$  \\ \hline
$\chi$ & 100 & $+0.440 \pm 0.019 $ & $+0.180 \pm 0.032$ & $+0.354 \pm 0.039$  \\ \hline 
$\eta$ & 100 & $+0.593 \pm 0.016 $ & $+0.257 \pm 0.033$ & $+0.561 \pm 0.039$  \\ \hline 
$\psi$ & 100 & $+0.673 \pm 0.012 $ & $+0.294 \pm 0.033$ & $+0.568 \pm 0.039$  \\ \hline 
$LR$   & 100 & $+0.303 \pm 0.022 $ & $+0.189 \pm 0.033$ & $+0.327 \pm 0.040$  \\ \hline 
\end{tabular}}
\label{tbl4}
\end{table}

There are, of course, other observables that one may try to use in the dilepton channel but they are somewhat more 
subtle. The first possibility{\cite {Dittmar:2003ir} is 
to reconstruct the Z' rapidity distribution from the dilepton final state. 
The left panel of Fig.~\ref{fig12} reminds us that the Z' rapidity distribution produced by only $u\bar u$, $d\bar d$ or 
sea quarks would have a different shape. The particular Z' couplings to quarks induce different weights in 
these three distributions and so one may hope to distinguish models in this way. An example of this is shown in the 
right panel of Fig.~\ref{fig12}. The first analysis{\cite {Dittmar:2003ir}} of this type 
considered the quantity $R_{u\bar u}$, the fraction of Z' events originating from $u\bar u$, as an observable; a similar 
variable $R_{d\bar d}$ can also be constructed. Fig~\ref{fig678} from a preliminary ATLAS analysis{\cite {Morel}}  
compares the values 
of these two parameters extracted via full reconstruction for a 1.5 TeV Z'; here we see that reasonable agreement with the 
input values of the generator are obtained although the statistical power is not very good. Knowing both 
$R_{d\bar d,u\bar u}$ and the ratio of the $d\bar d$ and $u\bar u$ parton densities fairly precisely, one can turn these 
measurements into a determination of the coupling ratio $(v_u^{'2}+a_u^{'2})/(v_d^{'2}+a_d^{'2})$.  
\begin{figure}
\centerline{\psfig{file=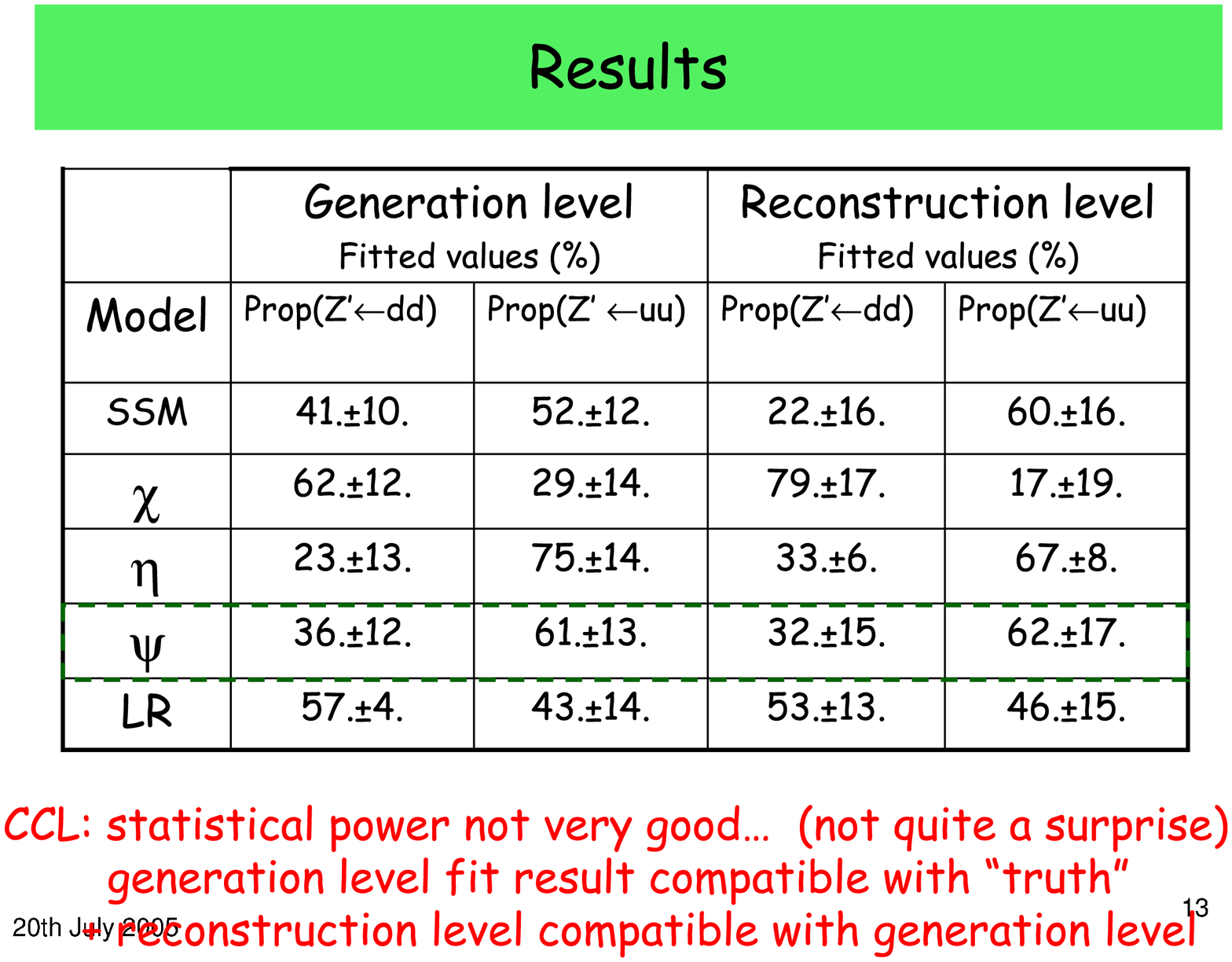,width=5.0cm,clip=,angle=-90}}
\caption{Comparison of $R_{q\bar q}$ values determined at the generator level and after detector simulation by ATLAS.}
\label{fig678}
\end{figure}

A second possibility is to construct the rapidity ratio{\cite {delAguila:1993ym}}} in the region near the Z' pole: 
\begin{equation}
R={{\int^{y_1}_{-y_1}~{{d\sigma} \over {dy}}~dy}\over{\Big[\int^{Y}_{y_1}+\int^{-y_1}_{-Y}~{{d\sigma} \over {dy}}
~dy\Big]}} 
\,.
\end{equation}
Here $y_1$ is some suitable chosen rapidity value $\simeq 1$. $R$ essentially measures the ratio of the cross section in the 
central region to that in the forward region and is again sensitive to the ratio of $u$ and $d$ quark couplings 
to the Z'. A detector level study of this observable has yet to be performed.
\begin{figure}
\centerline{\psfig{file=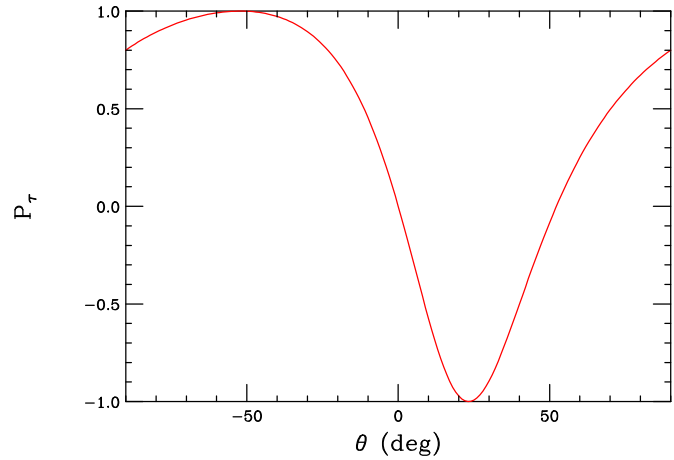,width=6.0cm,angle=90}}
\caption{$\tau$ polarization asymmetry for a Z' in $E_6$ models in the NWA.}
\label{fig12a}
\end{figure}

In addition to the $e^+e^-$ and $\mu^+\mu^-$ discovery channel final states, one might also consider other 
possibilities, the simplest being $\tau^+\tau^-$. Assuming universality, this channel does not provide anything new 
unless one can measure the polarization of the $\tau$'s, $P_\tau$, on or very near the Z' peak{\cite {Anderson:1992pc}}. 
The statistics for making this measurement can be rather good as the rate for this process is only smaller than that 
of the discovery mode by the $\tau$ pair reconstruction efficiency. In the NWA,  
$P_\tau=2v_e'a_e'/(v_e^{'2}+a_e^{'2})$, assuming universality, so that the ratio of $v_e'/a_e'$ can be determined 
uniquely. Fig.~\ref{fig12a} shows, for purposes of demonstration, the value of $P_\tau$ in the $E_6$ model case where 
we see that it covers its fully allowed range. 

A first pass theoretical study{\cite {Anderson:1992pc}} suggests that $\delta P_\tau \simeq 1.5/\sqrt N$, with $N$ here 
being the number of reconstructed $\tau$ events. Even for a reconstruction efficiency of $3\%$, with $M_{Z'}$ not too 
large $\sim 1-1.5$ TeV, the high luminosity of the LHC should be able to tell us $P_\tau$ at the $\pm 0.05$ level. 
It would be very good to see a detector study for this observable in the near future to see how well the LHC can 
really do in this case. 
\begin{figure}[ht]
\begin{center}
  \minifig{2.2in}{\epsfig{figure=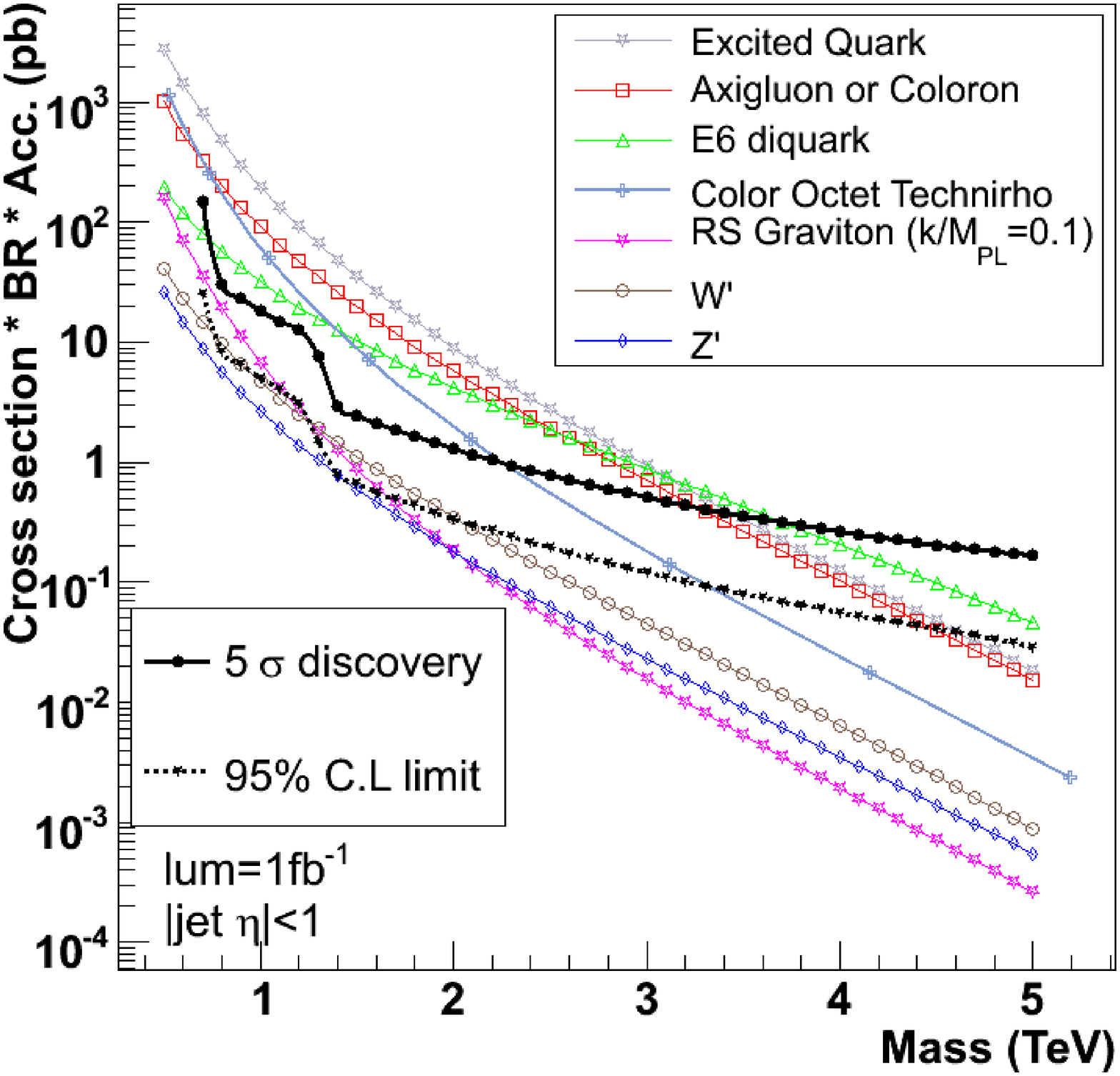,width=2.6in}}{a}
  \hspace*{4pt}
  \minifig{2.1in}{\epsfig{figure=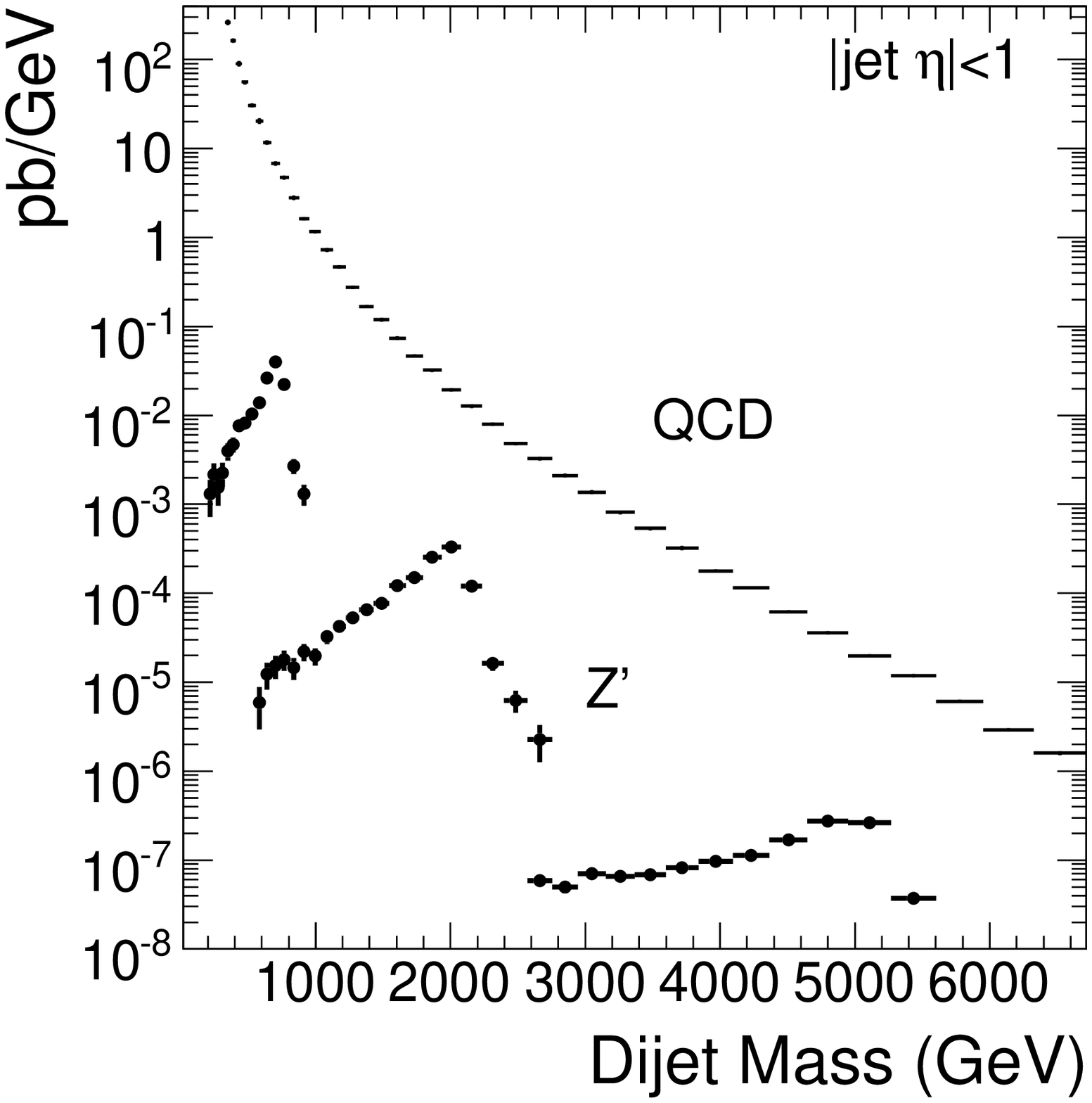,width=2.6in}}{b}
  \caption{(a)Dijet resonance discovery reach at CMS in comparison to the predictions for a number of 
models. (b) SSM Z' dijet signal for various masses in comparison with the SM background.}%
  \label{fig13}%
\end{center}
\end{figure}

Once we go beyond the dileptons, the next possibility one can imagine is light quark jets from which one might hope 
to get a handle on the Z' couplings to quarks. The possibility of new physics producing an observable dijet peak at the 
LHC has been studied in detail by CMS{\cite {Rob}}; the essential results are shown in Fig.~\ref{fig13}. Here we see that for 
resonances which are color non-singlets, \ie, those which have 
QCD-like couplings, the rates are sufficiently large as to allow these resonances to 
be seen above the dijet background. However, for weakly produced particles, such as the SSM Z' shown here, the 
backgrounds are far too large to allow observation of these decays. Thus it is very unlikely that the dijet channel will 
provide us with any information on Z' couplings at the LHC. 

Another possibility is to consider the heavy flavor decay modes, \ie, Z'$\to b\bar b$ or $t\bar t$. Unfortunately, these 
modes are difficult to observe so that it will be quite unlikely that we will obtain coupling information from them. 
ATLAS{\cite {Hoz}} has performed a study of the 
possibility of observing these modes within the Little Higgs Model context for a Z' in the 1-2 TeV mass range. 
Fig.~\ref{fig12aa} from the ATLAS study demonstrates how difficult observing these decays may really be due to the very 
large SM backgrounds. It is thus unlikely 
that these modes will provide any important information except in very special cases. 
\begin{figure}[ht]
\begin{center}
  \minifig{2.2in}{\epsfig{figure=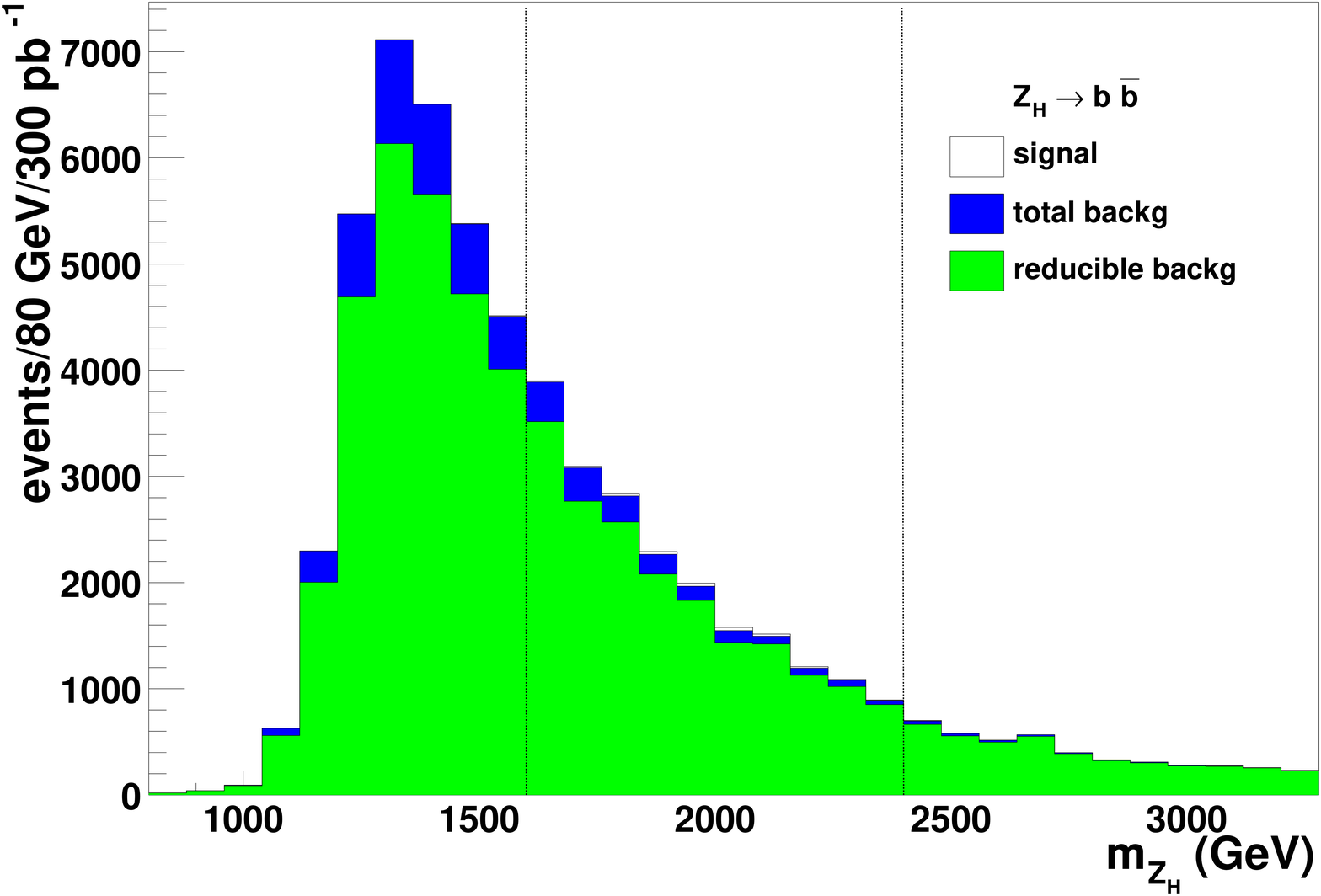,width=2.4in}}{a}
  \hspace*{4pt}
  \minifig{2.1in}{\epsfig{figure=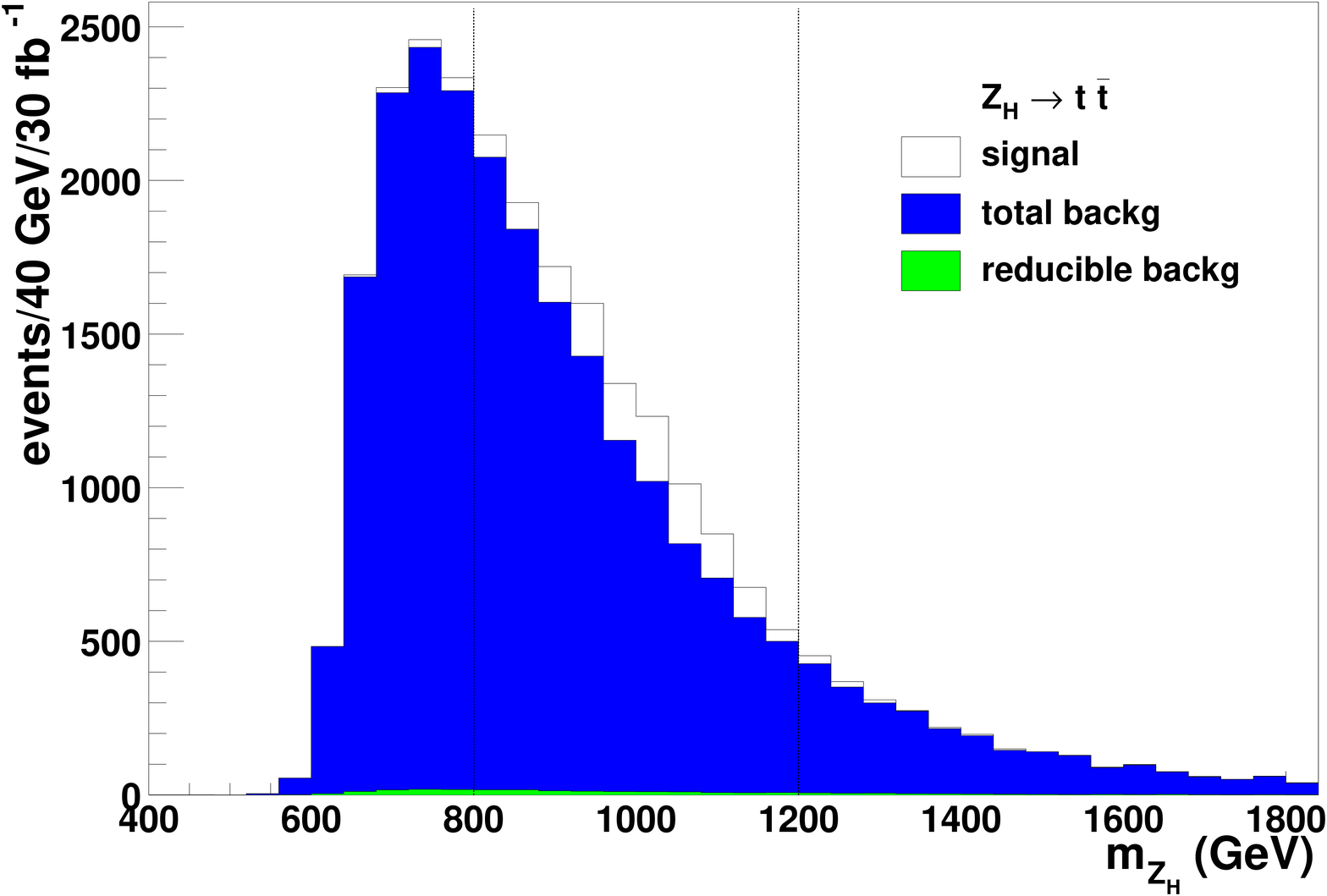,width=2.4in}}{b}
  \caption{Search for heavy flavor decays of the Z' in the Little Higgs model by ATLAS. $\cot \theta_H=1$ has been assumed. 
$Z'\to b\bar b$ assuming $M_{Z'}=2$ TeV and a luminosity of 300 $fb^{-1}$(a) and $t\bar t$(b) for $M_{Z'}=1$ TeV and a 
luminosity of 30 $fb^{-1}$.}%
  \label{fig12aa}%
\end{center}
\end{figure}

Another possible 2-body channel is Z'$\to W^+W^-$, which can occur at a reasonable rate through Z-Z' mixing as discussed 
above. Clearly the rate for this mode is very highly model dependent. ATLAS{\cite {Benchekroun}} has 
made a preliminary analysis 
of this mode in the $jjl\nu$ final state taking the Z' to be that of the SSM(for its fermionic couplings) and assuming 
a large integrated luminosity of 300 $fb^{-1}$. The mixing parameter $\beta$ was taken to be unity in the calculations. 
The authors of this analysis found that a Z' in the mass range below $\simeq 2.2$ TeV could be observed in this channel 
given these assumptions. An example is shown in Fig.~\ref{fig12abc} where we clearly see the reconstructed Z' above the 
SM background. With a full detailed background study an estimate could likely be made of the relevant branching 
fraction in comparison to that of the discovery mode. This would give important information on the nature of the Z' coupling 
structure. More study of this mode is needed. 
\begin{figure}
\centerline{\psfig{file=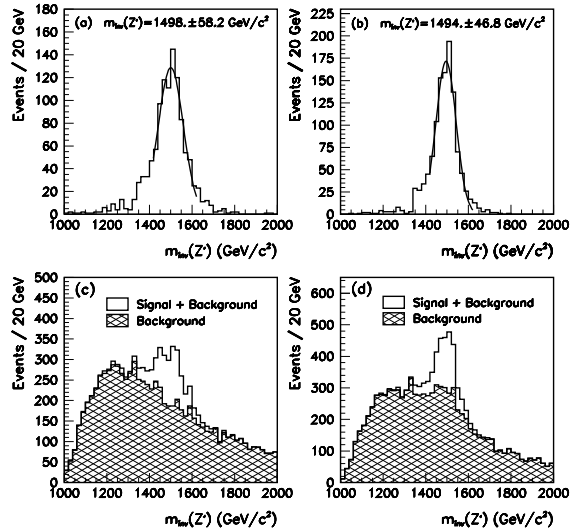,width=8.0cm,clip=,angle=0}}
\caption{Results of two ATLAS analyses showing the Z'$\to$ WW signal above SM backgrounds and Z' mass reconstruction 
in this channel for the SSM model assuming $M_{Z'}=1.5$ TeV and $\beta=1$.}
\label{fig12abc}
\end{figure}

A parallel study was performed by ATLAS{\cite{Azuelos:2004dm}} for the Z'$\to$ ZH mode which also occurs through mixing as 
discussed above; this mixing occurs naturally in the Little Higgs model in the absence of T-parity. 
The results are shown in Fig.~\ref{fig12aaa}. Here we see that there is a respectable signal over background 
and the relevant coupling information should be obtainable provided the Z' is not too heavy.
\begin{figure}[ht]
\begin{center}
  \minifig{2.2in}{\epsfig{figure=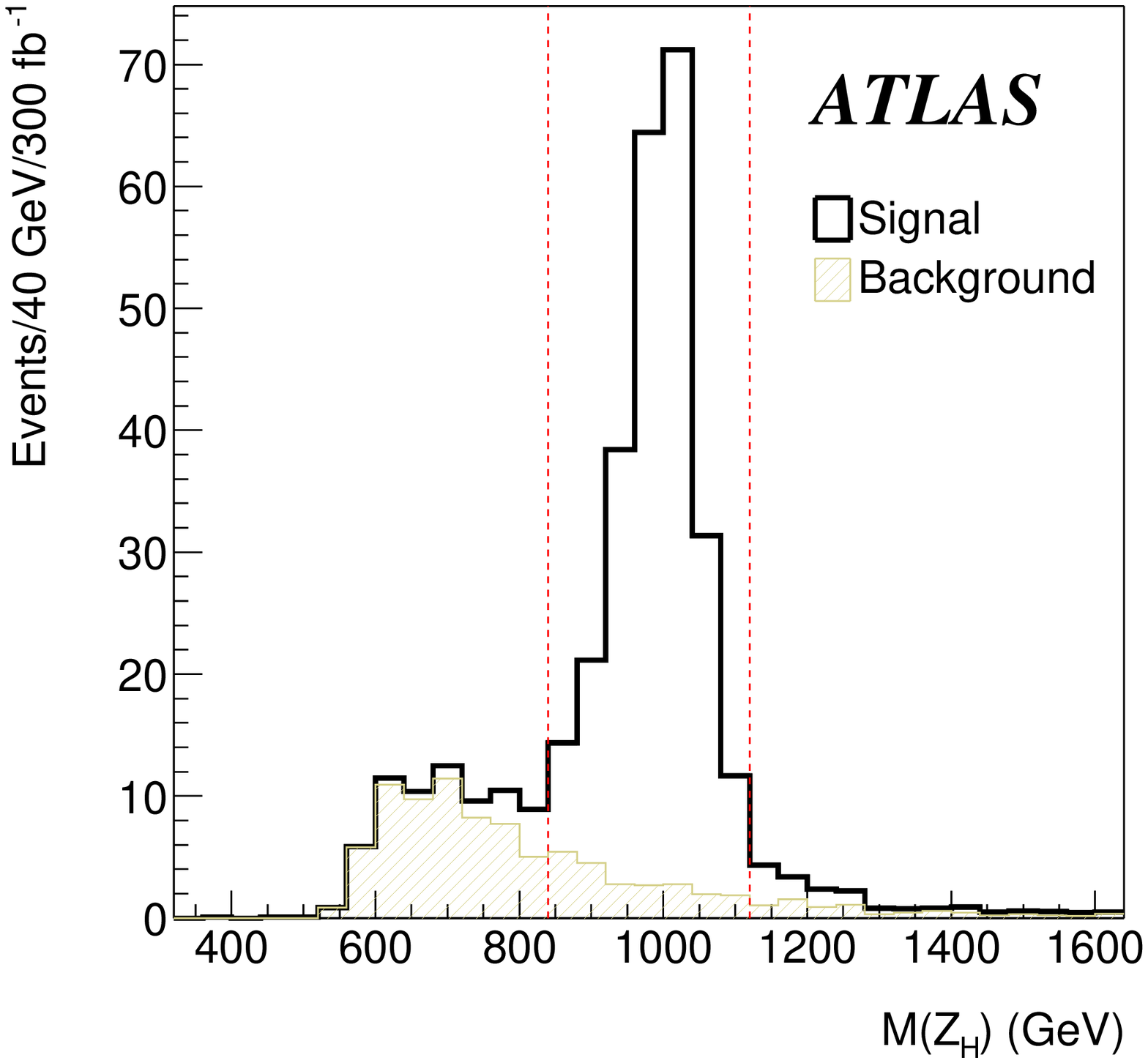,width=2.4in}}{a}
  \hspace*{4pt}
  \minifig{2.1in}{\epsfig{figure=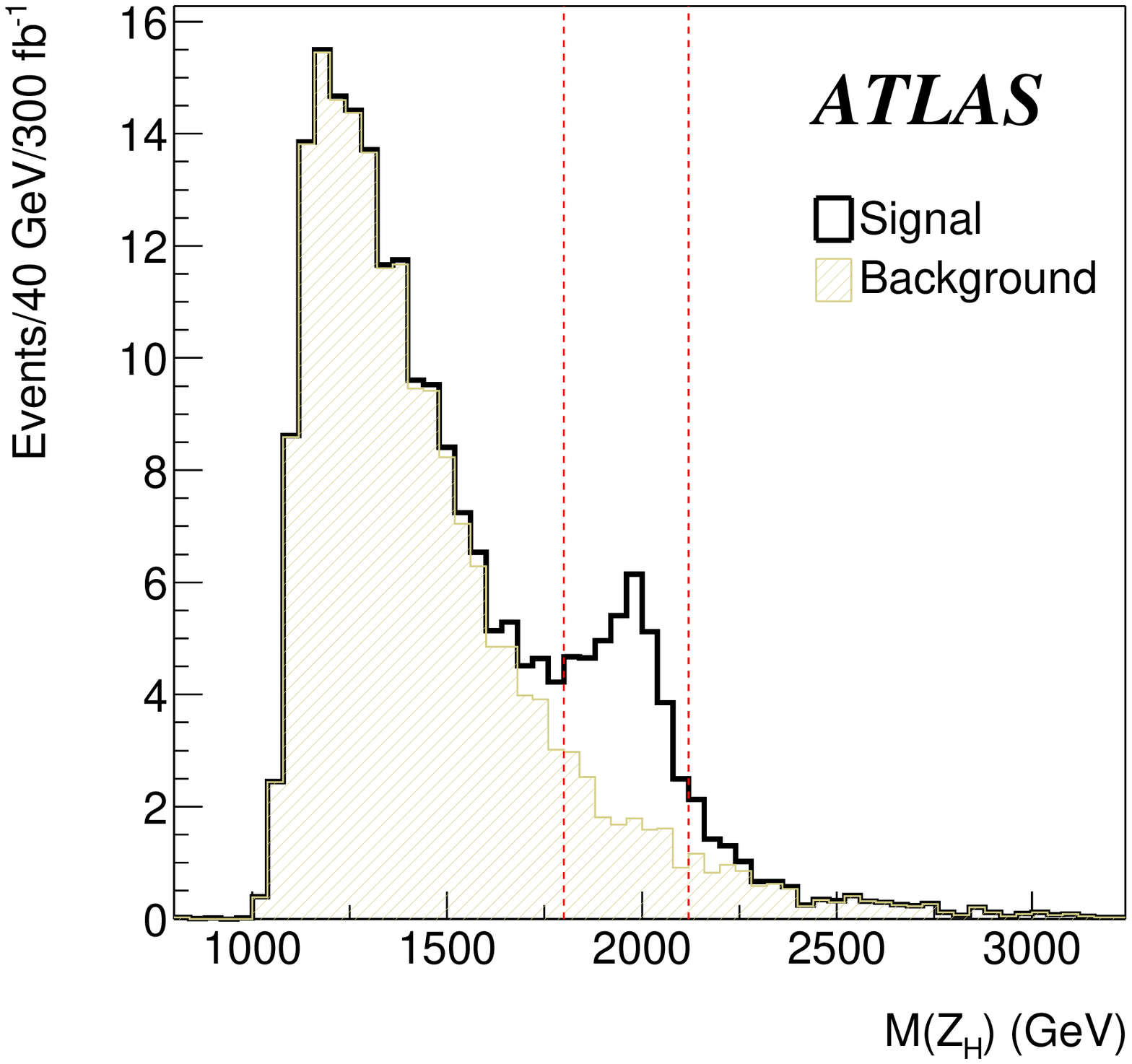,width=2.4in}}{b}
  \caption{Search study for the decay $Z' \to ZH$ by ATLAS in the Little Higgs model assuming $\cot \theta_H=0.5$ for the 
$l^+l^- b\bar b$ mode assuming $M_{Z'}$=1 (a) or 2(b) TeV.}%
  \label{fig12aaa}%
\end{center}
\end{figure}

Some rare decays of the Z' may be useful in obtaining coupling information provided the Z' is not too massive. Consider 
the ratios of Z' partial widths{\cite {Rizzo:1987rw,Cvetic:1991gk,delAguila:1993ym,Hewett:1992nf}}
\begin{equation}
r_{ff'V}={{\Gamma(Z'\to ff'V)}\over {\Gamma(Z'\to l^+l^-)}}\,,
\end{equation}
where $V=$Z,W and $ff'=l^+l^-,l^\pm \nu,\nu \bar \nu$, appropriately. The two $\Gamma(Z'\to f\bar fZ)$ (with $f=l,\nu$) 
partial widths originate from the bremsstrahlung of a SM Z off of either the $f$ or $\bar f$ legs and are rather 
to imagine. Numerically, one finds that for the case $f=l$, little sensitivity to the Z' couplings 
is obtained so it is not usually considered. Assuming that the SM $\nu$'s couple in a left-handed way to the Z', it is clear 
that $r_{\nu\nu Z}=K_Z v_\nu^{'2}/(v_e^{'2}+a_e^{'2})$, where $K_Z$ is a constant, model-independent factor for any given 
Z' mass. The signal for this decay is a (reconstructed) Z plus missing $p_T$ with a Jacobean peak at the Z' mass. 

$r_{l\nu W}$, on the 
otherhand, is more interesting; not only can the W be produced as a brem but it can also arise directly 
if a WWZ' coupling exists. 
As we saw above this can happen if Z-Z' mixing occurs {\it or} it can happen if $T'$ is proportional to $T_{3L}$. 
If there is no mixing and if $T'$ has no $T_{3L}$ component then one finds the simple relation $r_{l\nu W}=K_W 
v_\nu^{'2}/(v_e^{'2}+a_e^{'2})$, with $K_W$ another constant factor. Note that now $r_{l\nu W}$ and  $r_{\nu\nu Z}$ are 
proportional to one another and, since $T'$ and $T_{3L}$ commute, one also has $v_e'+a_e'=v_\nu'+a_\nu'=2v_\nu'$ so that 
both $r_{l\nu W}$ and  $r_{\nu\nu Z}$ are {\it bounded}, \ie,  $0\leq r_{l\nu W}\leq K_W/2$ and $0\leq r_{\nu\nu Z}\leq 
K_Z/2$.  Thus, \eg, in $E_6$ models a short analysis shows that the allowed region in the $r_{l\nu W},r_{\nu\nu Z}$ plane 
will be a straight line beginning at the origin and ending at $K_W/2,K_Z/2$. Other common models will lie on this line, such
as the LRM and ALRM cases, but some others, \eg, the SSM, will lie elsewhere in this plane 
signaling the fact that $T'$ contains a $T_{3L}$ component. Fig.~\ref{fig13a} from{\cite {Hewett:1992nf}} shows a 
plot of these parameters for a large number of models, the solid line being 
the just discussed $E_6$ case and `S' the SSM result. 

While the coupling information provided by these ratios is very 
useful, the Z' event rates necessary to extract them are quite high in most cases due to their small relative branching 
fractions. For a Z' much more massive than 1-2 TeV the statistical power of these observables will be lost. 
\begin{figure}
\centerline{\psfig{file=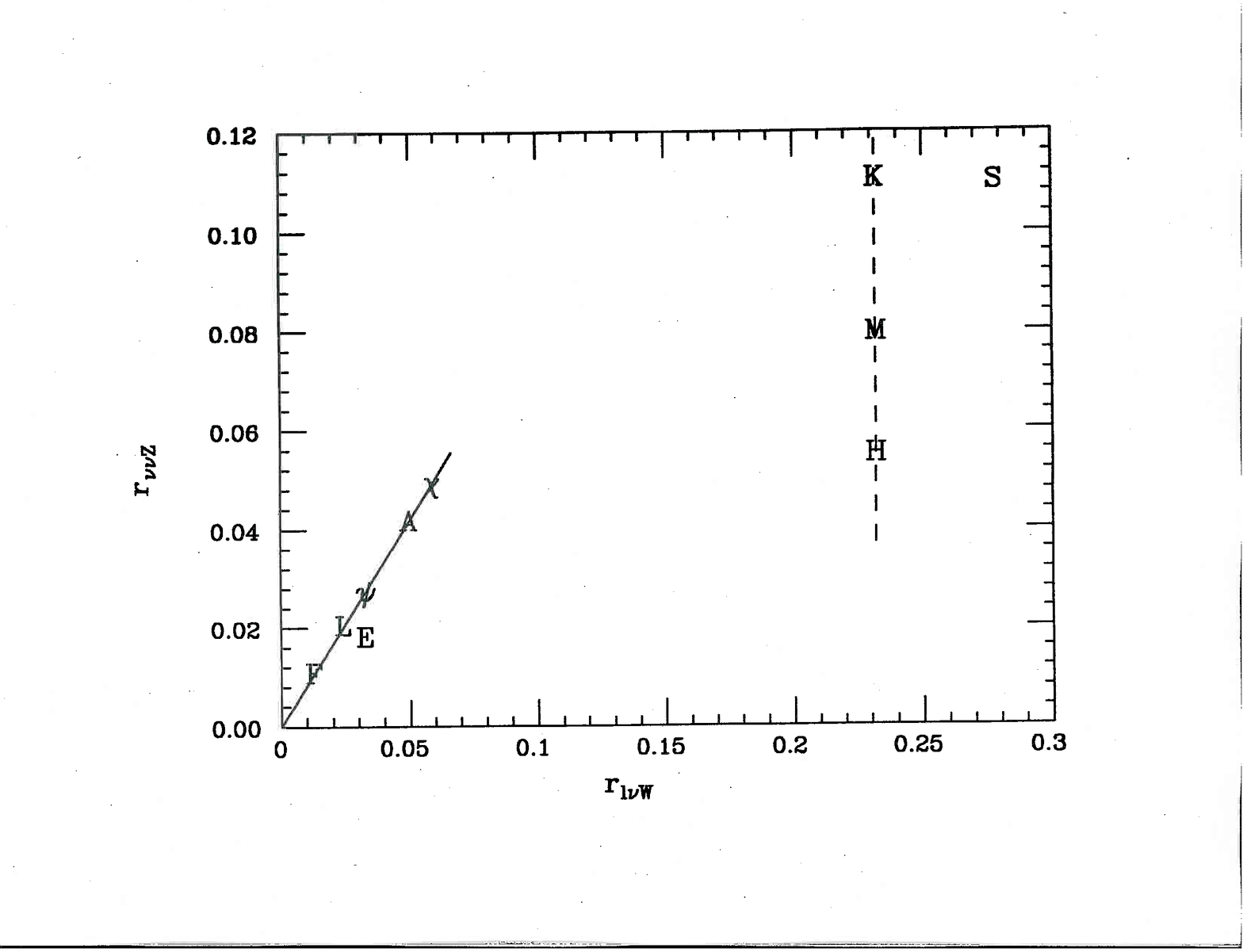,width=8.8cm,clip=,angle=-90}}
\caption{Predictions for the rare decay mode ratios for a number of different models assuming a 1 TeV Z': `L' is the 
LRM with $\kappa=1$, `S'=SSM, `A'=ALRM, \etc. The solid line is the $E_6$ case.}
\label{fig13a}
\end{figure}

A different way to get at the Z' couplings is to produce it in association with another SM gauge boson, \ie, a 
photon{\cite {Rizzo:1992sh}} or a W$^\pm$,Z{\cite {Cvetic:1992qv}}, with the Z' decaying 
to dileptons as usual. Taking the ratio of this cross section to that in the discovery channel, we can define the ratios 
\begin{equation}
R_{Z'V}={{\sigma(q\bar q \to Z'V)B(Z'\to l^+l^-)}\over {\sigma(q\bar q \to Z')B(Z'\to l^+l^-)}}\,,
\end{equation}
in the NWA with $V=\gamma$,W$^{\pm}$, or Z. (For the case $V=g$ there is little coupling sensitivity{\cite {Rizzo:1992sh}}). 
Note that $B$ trivially cancels in this ratio but it remains important for determining statistics. 
The appearance of an extra particle 
$V$ in the final state re-weights the combination of couplings which appears in the cross section so that one can get a 
handle on the vector and axial-vector couplings of the initial $u$'s and $d$'s to the Z'. For example, in 
the simple case of $V=\gamma$, the associated parton level $q\bar q \to Z'\gamma$ cross section is proportional to 
$\sum_i Q_i^2(v_i^{'2}+a_i^{'2})$ while the simple Z' cross section is proportional to $\sum_i (v_i^{'2}+a_i^{'2})$. 
Similarly, for the case $V=$W, the cross section is found to be proportional to $\sum_i (v_i'+a_i')^2$.   
Tagging the additional $V$, when $V\neq \gamma$, may require paying the price of leptonic branching 
fractions for the W and Z, which is a substantial rate penalty, although an analysis has not yet been performed. 
For the case of $V=\gamma$, a hard $p_T$ cut on the $\gamma$ 
will be required but otherwise the signature is very clean. All the ratios $R_{Z'V}$ are of order a 
few $\times 10^{-3}$ (or smaller once branching fractions are included) for a Z' mass of 1 TeV and (with fixed 
cuts) tend to grow with increasing $M_{Z'}$. For example, for a 1 TeV Z' in the $E_6$ model, the cross section times 
leptonic branching fraction for the Z'$\gamma$ final state varies in the range 0.65-1.6 fb, depending upon the parameter 
$\theta$, assuming a photon $p_T$ cut of 50 GeV. $R_\gamma$ for this case is shown in Fig.~\ref{fig14}. 
Generically, with 100 $fb^{-1}$ of luminosity these ratios might be determined at the  
level of $\simeq 10\%$ for the $M_{Z'}$=1 TeV case but the quality of the measurement will fall rapidly as $M_{Z'}$ increased 
due to quickly falling statistics. For much larger masses these ratios are no longer useful. It is possible that the Tevatron 
will tell us whether such light masses are already excluded. 
\begin{figure}
\centerline{\psfig{file=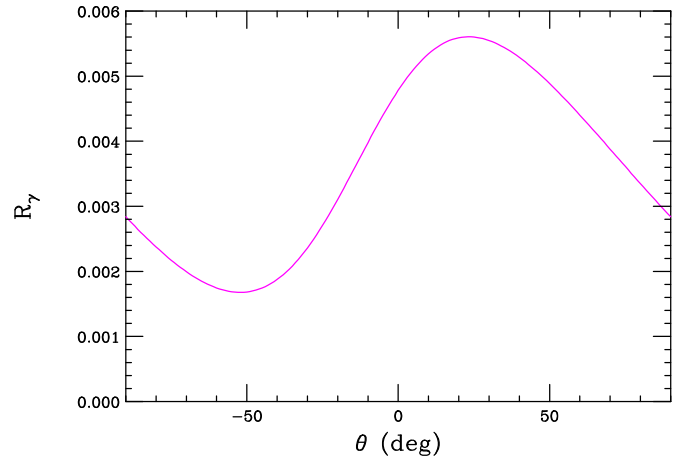,width=6.0cm,angle=90}}
\caption{$R_\gamma$ in $E_6$ models for a 1 TeV Z' employing a cut $p_T^\gamma >50$ GeV.}
\label{fig14}
\end{figure}

It is clear from the above discussion that there are many tools available at the LHC for Z' identification. However, many of 
these are only applicable if the Z' is relatively light. Even if all these observables are available it still 
remains unclear as to whether or not the complete set of Z' couplings can be extracted from the data with any reliability. 
A detailed analysis of this situation has yet to be performed. We will probably need a Z' discovery before it is done.

\section{ILC: What Comes Next}

The ILC will begin running a decade or so after the turn on of the LHC. At that point perhaps as much as $\sim 1~ab^{-1}$ or 
more of integrated luminosity will have been delivered by the LHC to both detectors. From our point of view, the role of 
the ILC would then be to either extend the Z' search reach (in an indirect manner) beyond that of the LHC or to help 
identify any Z' discovered at the LHC{\cite{Weiglein:2004hn}}. 

Although the ILC 
will run at $\sqrt s=0.5-1$ TeV, we know from our discussion of LEP Z' searches that the ILC will be sensitive to Z' 
with masses significantly larger than $\sqrt s$. Fig.~\ref{fig15}{\cite {Rizzo:2003sz}} shows the search reach for 
various Z' models assuming $\sqrt s=0.5,1$ TeV as a function of the integrated luminosity both with and without 
positron beam polarization. Recall that the various final states $e^+e^-\to f\bar f$, $f=e,\mu,\tau,c,b,t$ can all 
be used simultaneously to obtain high Z' mass  
sensitivity. The essential observables employed here are $d\sigma/dz$ and $A_{LR}(z)$, which is now available since the 
$e^-$ beam is at least $80\%$ polarized. One can also measure the polarization of $\tau$'s in the final state. 
This figure shows that the ILC will be sensitive to Z' masses in the range 
$(7-14) \sqrt s$ after a couple of years of design luminosity, the exact value depending on the particular Z' model. Thus we 
see that it it relatively easy at the ILC to extend the Z' reach beyond the 5-6 TeV value anticipated at the LHC. 
\begin{figure}[ht]
\begin{center}
  \minifig{2.2in}{\epsfig{figure=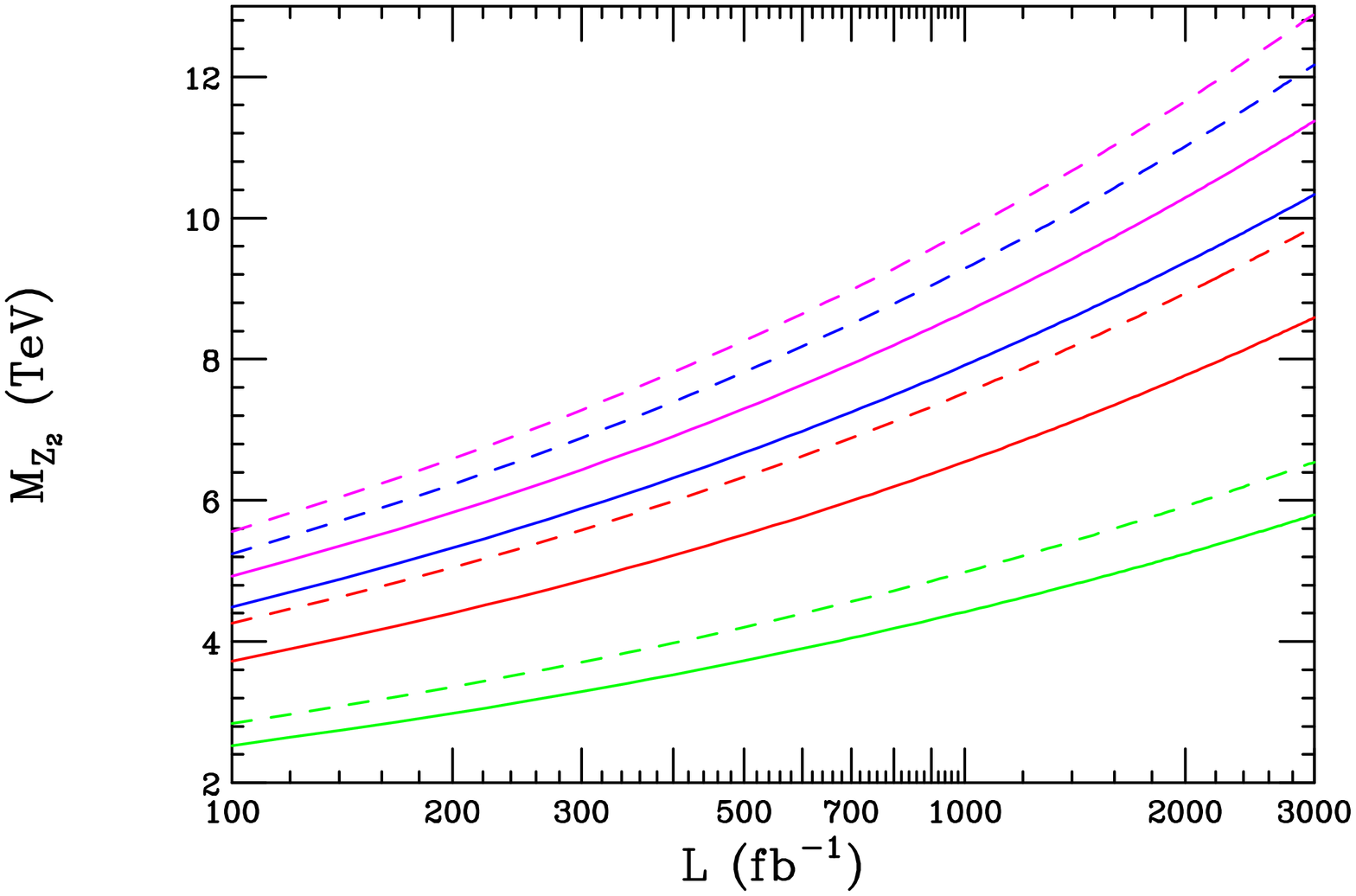,width=1.5in,angle=90}}{a}
  \hspace*{4pt}
  \minifig{2.1in}{\epsfig{figure=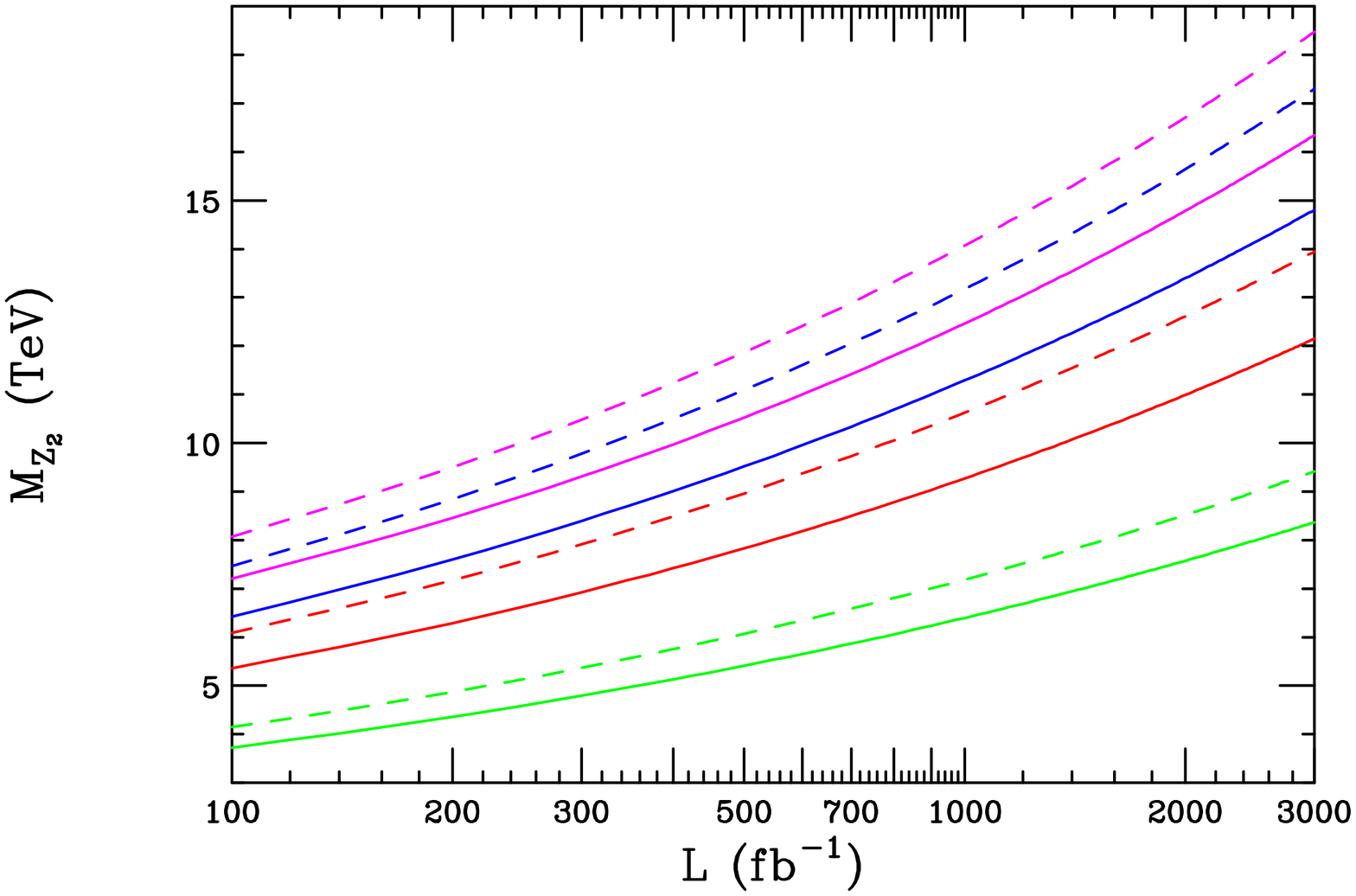,width=1.5in,angle=90}}{b}
  \caption{Z' search reach at a $\sqrt s$=0.5 TeV(a) or 1 TeV(b) ILC as a function of the 
integrated luminosity without(solid) 
or with(dashed) $60\%$ positron beam polarization for models $\psi$(green), $\chi$(red), SSM(magenta) and LRM with 
$\kappa=1$(blue).}%
  \label{fig15}%
\end{center}
\end{figure}
Fig.~\ref{fig15b} from {\cite {Godfrey}} shows a comparison of the direct Z' search reach at the LHC with the indirect reach 
at the ILC; note the very modest values assumed here for the ILC integrated luminosities. Here we see explicitly that the 
ILC has indirect Z' sensitivity beyond the direct reach of the LHC. 
\begin{figure}
\centerline{\psfig{file=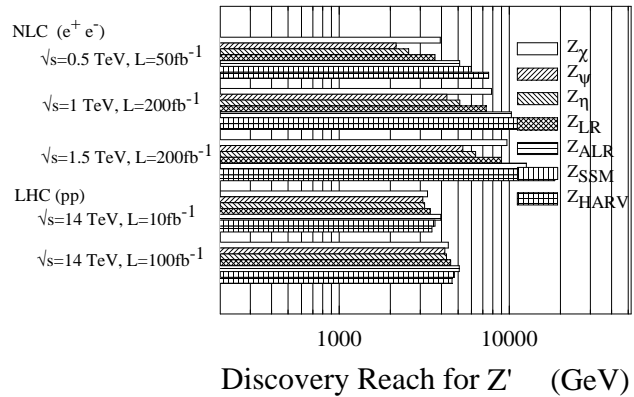,width=9.0cm,clip=}}
\caption{A comparison of LHC direct and ILC indirect Z' search reaches.}
\label{fig15b}
\end{figure}

In the more optimistic situation where a Z' is discovered at the LHC, the ILC will be essential for Z' identification. As 
discussed above, it is unclear whether or not the LHC can fully determine the Z' couplings, especially if it were much 
more massive than $\simeq 1$ TeV. 

Once a Z' is discovered at the LHC and its mass is determined, we can use the observed deviations 
in both $d\sigma/dz$ and $A_{LR}(z)$ 
at the ILC to determine the Z' couplings channel by channel. For example, assuming lepton universality (which we will 
already know is applicable from LHC data), we can examine the processes $e^+e^- \to l^+l^-$ using $M_{Z'}$ as an input  
and determine both $v_e'$ and $a_e'$ (up to a two-fold overall sign ambiguity); a measurement of $\tau$ polarization can 
also contribute in this channel. With this knowledge, we can go on to the $e^+e^- \to b\bar b$ channel and perform a 
simultaneous fit to $v_{e,b}'$ and $a_{e,b}'$; we could then go on to other channels such as $c\bar c$ and $t\bar t$. In 
this way {\it all} of the Z' couplings would be determined. An example of this is shown in Fig.~\ref{fig890} 
from {\cite {Riemann}} where we see the results of the Z' coupling determinations at the ILC in comparison 
with the predictions of a number of different models. 
\begin{figure}
\centerline{\psfig{file=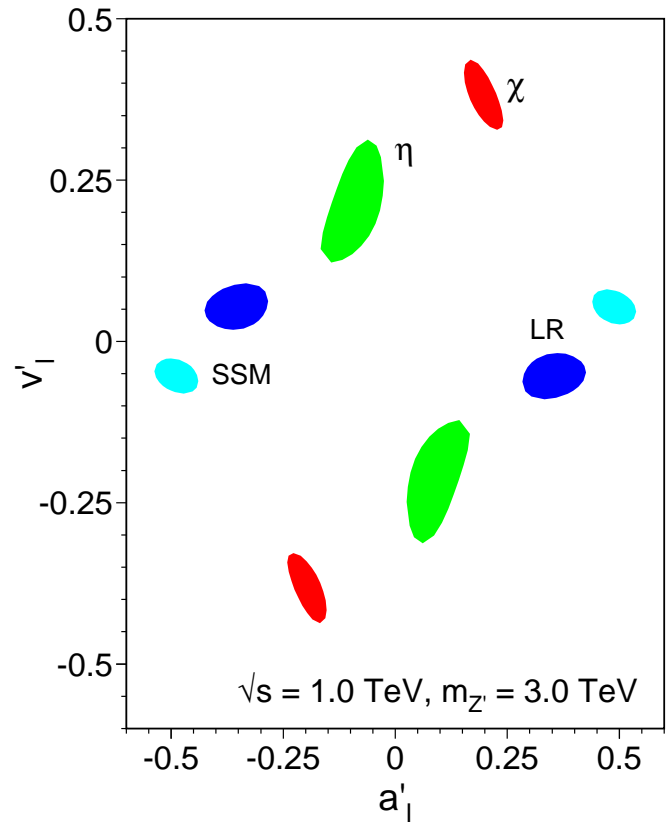,width=9.0cm}}
\caption{The ability of the ILC to determine the Z' leptonic couplings for a few representative models.}
\label{fig890}
\end{figure}

\section{Summary}

The LHC turns on at the end of next year and a reasonable integrated luminosity $\sim 1~fb^{-1}$ will likely be accumulated 
in 2008 at $\sqrt s=14$ TeV. The community-wide expectation is that new physics of some kind will be seen 
relatively `soon' after this (once the detectors 
are sufficiently well understood and SM backgrounds are correctly ascertained). Many new physics scenarios predict the 
existence of a Z' or Z'-like objects. It will then be up to the experimenters (with help from theorists!) to determine 
what these new states are and how they fit into a larger framework. In our discussion above, we have provided an overview 
of the tools which experiments at the LHC can employ to begin to address this problem. To complete this program will 
most likely require input from the ILC. 

No matter what new physics is discovered at the LHC the times ahead should prove to be very exciting.

\section*{Acknowledgments}

The  author would like to thank G. Azuelos, D. Benchekroun, C. Berger, K. Burkett, R. Cousins, A. De Roeck, S. Godfrey, 
R. Harris, J. Hewett, F. Ledroit, L. March, D. Rousseau, S. Willocq, and M. Woods for their input in the preparation of 
these brief lecture notes. Work supported in part by the Department of Energy, Contract DE-AC02-76SF00515.




\end{document}